\documentclass[final,3p,times,fleqn]{elsarticle}
\usepackage{mathtools,booktabs}
\usepackage{bm}
\usepackage{autobreak}
\usepackage{amsmath}
\usepackage{amsthm}
\usepackage{amssymb}
\usepackage{mathrsfs}
\usepackage{siunitx}
\usepackage{gensymb}
\usepackage{subfigure}
\usepackage{epstopdf}
\usepackage{color}
\usepackage{hyperref}
\usepackage{numprint}

\biboptions{sort&compress}
\journal{}
\begin{document}
	\begin{frontmatter}
		\title{Phase-field modeling of dendritic growth with gas bubbles in the solidification of binary alloys}
		\author[a,b,c]{Chengjie Zhan}
		\author[a,b,c]{Zhenhua Chai\corref{cor1}}
		\ead{hustczh@hust.edu.cn}
		\author[d]{Dongke Sun}
		\author[a,b,c]{Baochang Shi}
		\author[e]{Shaoning Geng}
		\author[e]{Ping Jiang}
		\address[a]{School of Mathematics and Statistics, Huazhong University of Science and Technology, Wuhan 430074, China}
		\address[b]{Institute of Interdisciplinary Research for Mathematics and Applied Science, Huazhong University of Science and Technology, Wuhan 430074, China}
		\address[c]{Hubei Key Laboratory of Engineering Modeling and Scientific Computing, Huazhong University of Science and Technology, Wuhan 430074, China}
		\address[d]{School of Mechanical Engineering, Southeast University, Nanjing 211189, China}
		\address[e]{The State Key Laboratory of Digital Manufacturing Equipment and Technology, School of Mechanical Science and Engineering, Huazhong University of Science and Technology, Wuhan 430074, China}
		\cortext[cor1]{Corresponding author.}
		\begin{abstract} 
			In this work, a phase-field model is developed for the dendritic growth with gas bubbles in the solidification of binary alloys. In this model, a total free energy for the complex gas-liquid-dendrite system is proposed through considering the interactions of gas bubbles, liquid melt and solid dendrites, and it can reduce to the energy for gas-liquid flows in the region far from the solid phase, while degenerate to the energy for thermosolutal dendritic growth when the gas bubble disappears. The governing equations are usually obtained by minimizing the total free energy, but here some modifications are made to improve the capacity of the conservative phase-field equation for gas bubbles and convection-diffusion equation for solute transfer. Additionally, through the asymptotic analysis of the thin-interface limit, the present general phase-field model for alloy solidification can match the corresponding free boundary problem, and it is identical to the commonly used models under a specific choice of model parameters. Furthermore, to describe the fluid flow, the incompressible Navier-Stokes equations are adopted in the entire domain including gas, liquid, and solid regions, where the fluid-structure interaction is considered by a simple diffuse-interface method. To test the present phase-field model, the lattice Boltzmann method is used to study several problems of gas-liquid flows, dendritic growth as well as the solidification in presence of gas bubbles, and a good performance of the present model for such complex problems is observed. 	    
		\end{abstract}
	\end{frontmatter}		
\section{Introduction}
The microstructural evolution during the process of solidification usually involves the formation of the gas bubbles due to the solubility difference of gas phase in liquid and solid phases as well as the gas entrapment during the melt filling \cite{Lee2001JLM}. Under the effect of the gravity, the bubbles rise slowly in the liquid metal with a high viscosity, and are more likely to be trapped during the transient solidification. This would cause the formation of porosity defect and have an influence on the mechanical properties, such as the tensile strength and fatigue resistance \cite{Ahmad2005JCM,Irfan2012MSEA}. 
To clarify the evolution mechanism of bubbles and improve the mechanical properties of materials, several different kinds of approaches, including theoretical, experimental and numerical methods, have been developed to understand the formation and evolution of the gas bubble during the solidification. Theoretical methods are useful in predicting the position and size of the microporosity \cite{Sigworth1993MTB,Han2002MMTA}, but they cannot determine the specific morphology of the microstructure. Experimental methods can provide the valuable insight in verifying the theory of pore formation and actual phenomena in the solidifying materials \cite{Lee1997AM,Xing2010AM,Murphy2016MSE,Bhagavath2019MMTA}. However, it is difficult to carry out some quantitative experimental researches for the solidification due to the limitations of measurement techniques and influences of many physical factors and processes. 

With the development of the computer science and scientific computing, the numerical simulation has become an alternative tool to investigate the mechanism behind the bubble deformation during the solidification and the interaction between solid dendrites and gas bubbles \cite{Atwood2002MMTB,Meidani2011AM,Karagadde2012CMS}. However, the solidification problem consists of different physical fields that need to be solved simultaneously, which brings great challenges in modeling and simulation of such a complex system. 
To describe the evolution of the liquid-solid interface in the solidification, some mathematical models and methods have been developed \cite{Szep1985JPA,Beckermann1999JCP,Tan2007JCP,Voller2008IJHMT}, and among them, the cellular automation (CA) \cite{Zhu2007AM,Luo2013CMS} and phase-field method (PFM) \cite{Boettinger2002ARMR,Chen2002ARMR} are the most popular ones. Actually, the CA is efficient but not accurate enough, the PFM can simulate the realistic dendrite patterns, but the cost is relatively expensive due to the computation of the order parameter in the entire physical domain. 
On the other hand, the gas-liquid two-phase flows are ubiquitous in nature and daily life, and have been well modeled by some sharp-interface approaches \cite{Hou2001JCP,Unverdi1992JCP,Sussman1994JCP} and diffuse-interface approaches \cite{Anderson1998ARFM,Jacqmin1999JCP,Sun2008PD}, in which the PFM \cite{Jacqmin1999JCP} is popular for its features of the needless explicit interface-tracking, the thermodynamic consistency, and the accuracy even for the problems with high density ratios. 
To investigate the solidification in presence of gas bubbles, the natural way is to directly couple above two kinds of methods.
In the past years, the CA coupled with the Shan-Chen lattice Boltzmann method (LBM) has been widely used to study the bubble growth and movement in the process of the solidification because of the mesoscopic modeling characteristics of these two methods \cite{Wu2012MSE,Sun2016IJHMT,Zhang2020IJHMT}. However, the Shan-Chen model usually suffers from the deficiency of the large spurious currents even though some improved models have been proposed \cite{Chen2014IJHMT}, which may affect the numerical accuracy and stability of the method. 
To produce more accurate results on the solidification process in presence of gas bubbles, the PFM has also been applied \cite{Nabavizadeh2019AS,Nabavizadeh2019IJMF}. However, the PFM is only adopted for the gas-liquid interface capturing, and another method is considered for the liquid-solid interface evolution, such as the coupled CA-PFM \cite{Nabavizadeh2019IJMF}. 

To develop a more efficient and accurate method in the unified diffuse-interface framework, a phase-field based LBM for the gas-liquid-solid interaction during the solidification was proposed in Ref. \cite{Zhang2020MMTA}. However, the computational region of each governing equation and the details on how to treat complex boundary conditions are not presented. Zhang et al. \cite{Zhang2021AM} further proposed another PFM to model the gas-liquid-solid interaction in a unified scheme, which is based on a multiphase-field concept \cite{Steinbach1996PDNP}. In this model, two energy function terms are added to adjust the height of the two-phase energy barrier and the height of the three-phase junction region in the free-energy landscape \cite{Folch2005PRE,Wang2018PRM}, and the Lagrange multiplier is also introduced to preserve the total conservation of the phase fraction. However, this model is not variational, and many step functions are introduced in their diffuse-interface PFM. Recently, Zhang et al. \cite{Zhang2022Desalination} investigate the effect of single bubble behavior on the seawater-frozen crystallization by using a coupled PFM-LBM. However, in this work, it is confusing that the gas-liquid interface is described by the Cahn-Hilliard equation while the interaction force is determined by the Shan-Chen model in LBM. Instead of directly designing the governing equations for such a complex system, Du et al. \cite{Du2016CMS} proposed a thermodynamically consistent PFM by establishing an underlying total free energy for the system of the gas bubble nucleation and growth in the solidification of pure metal, in which the interaction energy between phase parameter and gas concentration is well-defined, but the melt flow is not considered. 

In this paper, to describe the binary alloy solidification in presence of gas bubbles by a unified diffuse-interface method, we will develop a phase-field model where the interaction between dendrites and bubbles is described by introducing an interaction energy function in the total free energy. Based on the phase-field theory, we propose a total free energy for the gas-liquid-dendrite system, and the free energy functional would reduce to the classic energy for gas-liquid two-phase flows in the region far from the solid phase, while degenerate to the one for thermosolutal dendritic growth in the absence of gas bubbles. Through minimizing the present free energy functional, we can obtain the conservative phase-field equation for capturing gas-liquid interface, the non-conservative one for solidification, and the convection-diffusion equation for solute transfer. In additional, the equation of heat transfer can be derived from the internal energy transport with the thermodynamic requirement of a positive entropy production. It is worth noting that through the asymptotic analysis of the thin-interface limit, the present phase-field model for alloy solidification can match the corresponding free boundary problem. Moreover, the force caused by the fluid-solid interaction can be depicted by a simple diffuse-interface method, instead of a dissipative drag force with an empirical constant \cite{Beckermann1999JCP}. To solve the proposed phase-field model, the mesoscopic LBM is adopted due to its distinct features of simplicity in coding and fully parallel algorithm \cite{Higuera1989EL,Chen1998ARFM,Aidun2010ARFM}. 

The rest of this paper is organized as follows. In Section \ref{sec-model}, the phase-field equations for binary alloy solidification in presence of gas bubbles are derived from the proposed total free energy, and followed by incompressible Navier-Stokes equations for the fluid flows. Then the numerical method and validations are presented in Section \ref{sec-method}. In Section \ref{sec-results}, the numerical results and discussion are given, and finally, some conclusions are summarized in Section \ref{sec-conclusion}. 

\section{Mathematical model}\label{sec-model}
\subsection{Phase-field model}
The total free energy of the system with bubble dynamics and dendritic growth in a binary alloy melt can be designed by
\begin{equation}\label{eq-EnergyTotal}
	\mathcal{F}_{total}=\int_{V}\left[f\left(\phi,\psi,c,T\right)+\frac{W_{\phi}^2}{2}|\nabla\phi|^2+\frac{W_{\psi}^2}{2}|\nabla\psi|^2\right]\mathrm{d}V,
\end{equation}  
where $\phi$ is the phase-field variable introduced to identify the gas bubble ($\phi=\phi_g$) and other phases ($\phi=\phi_l$), and $\psi$ is the order parameter used to depict the interface of the solid dendrite ($\psi=\psi_s$) and other phases ($\psi=\psi_l$). $W_{\phi}$ and $W_{\psi}$ are two physical parameters related to the thickness of the gas-liquid and liquid-solid interfaces. The bulk free energy density $f\left(\phi,\psi,c,T\right)$ is composed of three parts,
\begin{equation}
	f\left(\phi,\psi,c,T\right)=f_1\left(\phi\right)+f_{AB}\left(\psi,c,T\right)+f_{3}\left(\phi,\psi\right),
\end{equation}
where $f_1\left(\phi\right)$ is the bulk free energy of gas and liquid phases, $f_{AB}\left(\psi,c,T\right)$ is that of a binary mixture of $A$ and $B$ molecules, $f_{3}\left(\phi,\psi\right)$ represents the interaction energy between solid dendrites and gas bubbles, $c$ and $T$ represent the solute concentration of $B$ and temperature, respectively.
 
The bulk free energy density $f_{AB}\left(\psi,c,T\right)$ is usually written as a summation of the free energy of the pure material $f_2\left(\psi,T\right)$ and the contribution due to the solute addition \cite{Ramirez2004PRE},
\begin{equation}
	f_{AB}\left(\psi,c,T\right)=f_2\left(\psi,T\right)+\frac{RT}{\nu_0}\left(c\ln c-c\right)+\varepsilon\left(\psi\right)c,
\end{equation}
where $R$ is the universal gas constant, $v_0$ is the molar volume assumed to be a constant, and $\varepsilon\left(\psi\right)$ is an interpolation function between $\varepsilon_s$ in solid and $\varepsilon_l$ in liquid. 
After expanding $f_2\left(\psi,T\right)$ to first order in $\Delta T=T-T_M$ by defining $f_T\left(\psi\right)=\partial f_2\left(\psi,T\right)/\partial T|_{T=T_M}$ and replacing $RT/v_0$ by $RT_M/v_0$ \cite{Ramirez2004PRE}, the free energy density of binary alloy can be approximated by 
\begin{equation}
	f_{AB}\left(\psi,c,T\right)=f_2\left(\psi,T_M\right)+f_T\left(\psi\right)\Delta T+\frac{RT_M}{\nu_0}\left(c\ln c-c\right)+\left[\varepsilon_l+\Delta\varepsilon \bar{p}\left(\psi\right)\right]c,
\end{equation}  
here $\Delta\varepsilon=\varepsilon_s-\varepsilon_l>0$, $\bar{p}\left(\psi\right)$ is a monotonously increasing function with $\bar{p}\left(\psi_l\right)=0$, $\bar{p}\left(\psi_s\right)=1$, $\bar{p}'\left(\psi_s\right)=\bar{p}'\left(\psi_l\right)=0$, and $\bar{p}''\left(\psi_s\right)=\bar{p}''\left(\psi_l\right)=0$. The bulk free energy densities of gas-liquid part and pure material part usually have the standard forms of double-well profile,
\begin{equation}\label{eq-douWell}
	f_1\left(\phi\right)=\beta_{\phi}\left(\phi_l-\phi\right)^2\left(\phi-\phi_g\right)^2,\quad
	f_2\left(\psi,T_M\right)=\beta_{\psi}\left(\psi_s-\psi\right)^2\left(\psi-\psi_l\right)^2,
\end{equation}
where $\beta_{\phi}$ and $\beta_{\psi}$ are two physical parameters, and in this work, it is assumed that $\phi_l>\phi_g$ and $\psi_s>\psi_l$.
In addition, inspired by the obstacle potential that penalizes fields for overlapping \cite{Chen1994PRB,Chen1995SMM}, we propose an interaction energy between solid dendrites and gas bubbles as
\begin{equation}\label{eq-f3Inter}
	f_3\left(\phi,\psi\right)=\beta_{inter}\left(\phi_l-\phi\right)^2\left(\psi-\psi_l\right)^2,
\end{equation}
where the parameter $\beta_{inter}$ is used to determine the interaction factor. 
Here one can find that the function $f_3\left(\phi,\psi\right)$ satisfies the requirement of no interactions in absence of the solid or gas phase, and it only works at the gas-solid interface.

The equilibrium properties of the present model satisfy the following conditions,
\begin{equation}\label{eq-Eq}
	\frac{\delta\mathcal{F}_{total}}{\delta\phi}=0,\quad\frac{\delta\mathcal{F}_{total}}{\delta c}=\mu_c,\quad\frac{\delta\mathcal{F}_{total}}{\delta\psi}=0,
\end{equation}
where $\mu_c$ is the spatially uniform equilibrium value of the chemical potential. These three conditions uniquely determine the stationary profiles of $\phi$, $\psi$ and $c$ in the diffuse-interface region, $\phi^{eq}$, $\psi^{eq}$ and $c^{eq}$, and the details will be shown below.
\begin{itemize}
	\item[(1)] In the region far from solid phase, i.e., $\psi=\psi_l$, the first condition in Eq. (\ref{eq-Eq}) leads to the following equilibrium profile in the gas-liquid diffuse-interface region,
	\begin{equation}\label{eq-phiEq}
		\phi^{eq}\left(x\right)=\frac{\phi_l+\phi_g}{2}+\frac{\phi_l-\phi_g}{2}\tanh\left[\frac{\left(\phi_l-\phi_g\right)\sqrt{2\beta_{\phi}}}{2W_{\phi}}x\right].
	\end{equation} 
	\item[(2)] In the solidification region, i.e., $\phi=\phi_l$, the profile of $c$ varies between the equilibrium concentration $c_s^{eq}$ in solid phase and $c_l^{eq}$ in liquid phase, which satisfies 
	\begin{equation}
		\frac{\partial f_{AB}\left(c,T\right)}{\partial c}|_{c=c_s^{eq}}=\frac{\partial f_{AB}\left(c,T\right)}{\partial c}|_{c=c_l^{eq}}=\mu_c.
	\end{equation}
	Applying the second condition in Eq. (\ref{eq-Eq}), we can obtain
	\begin{equation}
		\frac{RT_M}{\nu_0}\ln c+\varepsilon_l+\Delta\varepsilon\bar{p}\left(\psi\right)=\mu_c,
	\end{equation}
	from which the expressions of the equilibrium partition coefficient $k$ and the stationary concentration profile $c^{eq}$ can be given by
	\begin{equation}
		k=\frac{c_s^{eq}}{c_l^{eq}}=\exp\left(-\frac{v_0\Delta\varepsilon}{RT_M}\right),\quad c^{eq}\left(x\right)=c_l^{eq}\exp\left\{\bar{p}\left[\psi\left(x\right)\right]\ln k\right\}.
	\end{equation}
	\item[(3)] Applying the third condition in Eq. (\ref{eq-Eq}) yields
	\begin{equation}\label{eq-muPsi=0}
		f'_T\left(\psi^{eq}\right)\Delta T+\bar{p}'\left(\psi^{eq}\right)\Delta\varepsilon c^{eq}=W_{\psi}^2\frac{\mathrm{d}^2\psi^{eq}}{\mathrm{d}x^2}-f'_2\left(\psi^{eq},T_M\right).
	\end{equation}
	If the function $f_T\left(\psi\right)$ takes the following form,
	\begin{equation}\label{eq-f_T}
		f_T\left(\psi\right)=\frac{RT_M}{\nu_0m}\exp\left[\bar{p}\left(\psi\right)\ln k\right],
	\end{equation} 
	the term on the left-hand side of Eq. (\ref{eq-muPsi=0}) would vanish under the condition $T=T_M+mc_l^{eq}$ with $m$ being the slope of the liquidus line in the phase diagram, and the following equilibrium profile can be derived,
	\begin{equation}\label{eq-psiEq}
		\psi^{eq}\left(x\right)=\frac{\psi_s+\psi_l}{2}+\frac{\psi_s-\psi_l}{2}\tanh\left[\frac{\left(\psi_s-\psi_l\right)\sqrt{2\beta_{\psi}}}{2W_{\psi}}x\right].
	\end{equation}
	In addition, since $f_T\left(\psi_s\right)-f_T\left(\psi_l\right)=L/T_M$ \cite{Karma2001}, one can get the Van't Hoff relation from Eq. (\ref{eq-f_T}),
	\begin{equation}
		\frac{L}{T_M}=\frac{RT_M\left(k-1\right)}{mv_0},
	\end{equation} 
	where $L$ is the latent heat of fusion. 
\end{itemize}

According to above definition of the total free energy, the governing equations for binary alloy solidification in presence of gas bubbles can be given by
\begin{subequations}
	\begin{equation}\label{eq-CHE}
		\frac{\partial\phi}{\partial t}=\nabla\cdot M_{\phi}\nabla\frac{\delta\mathcal{F}_{total}}{\delta\phi},
	\end{equation}
	\begin{equation}\label{eq-psi}
		\tau\frac{\partial\psi}{\partial t}=-\frac{\delta\mathcal{F}_{total}}{\delta\psi},
	\end{equation}
	\begin{equation}\label{eq-c}
		\frac{\partial c}{\partial t}=\nabla\cdot\left( M_c\nabla\frac{\delta\mathcal{F}_{total}}{\delta c}-\bar{\mathbf{J}}_{at}-\bar{\mathbf{J}}\right),
	\end{equation}
	\begin{equation}\label{eq-tem}
		\frac{\partial}{\partial t}\left[\rho c_p\left(T-T_M\right)-\rho L\bar{p}\left(\psi\right)\right]=\nabla\cdot k_T\nabla T,
	\end{equation}
\end{subequations}
where $M_{\phi}$ is the mobility, $\tau$ is the kinetic characteristic time, $M_c=v_0D\bar{q}\left(\psi\right)c/RT_M$, and in the liquid melt, the Fick's law of diffusion with diffusivity $D$ can be reduced through the expression of $\bar{q}\left(\psi\right)$. $\bar{\mathbf{J}}_{at}$ is an antitrapping current term used to counterbalance spurious solute trapping without introducing other thin-interface effects \cite{Karma2001PRL}, and $\bar{\mathbf{J}}$ is a flux term used to preserve a local solute vacuum in the gas phase. The governing equation of the temperature (\ref{eq-tem}) is derived by the condition of positive entropy production from the internal energy transport \cite{Wang1993PD}, $\rho$ is the density, $c_p$ is the specific heat per unit volume, and $k_T$ is the thermal conductivity. 

Furthermore, the governing equations for solidification can be rewritten in terms of a dimensionless variable $u$,
\begin{equation}
	u=\frac{v_0}{RT_M}\left(\mu_c-\mu_{c_{\infty}}\right)=\ln\frac{c}{c_{\infty}}-\bar{p}\left(\psi\right)\ln k,
\end{equation}
where $c_{\infty}$ is the value of $c$ far from the interface that equals the initial concentration of the alloy. The function $\bar{p}\left(\psi\right)$ can be replaced by the function $p\left(\psi\right)$ through the relation $\exp\left[\bar{p}\left(\psi\right)\ln k\right]=1-\left(1-k\right)p\left(\psi\right)$ \cite{Ramirez2004PRE}, then one can obtain
\begin{equation}\label{eq-u}
	u=\ln\frac{c/c_{\infty}}{1-\left(1-k\right)p\left(\psi\right)},\quad c^{eq}=c_l^{eq}\left[1-\left(1-k\right)p\left(\psi\right)\right].
\end{equation} 
After some simple algebraic manipulations, the phase-field and concentration equations for solidification [Eqs. (\ref{eq-psi}) and (\ref{eq-c})] can be rewritten as
\begin{equation}
	\tau\frac{\partial\psi}{\partial t}=W_{\psi}^2\nabla^2\psi-f'_2\left(\psi,T_M\right)-\partial_{\psi}f_3\left(\phi,\psi\right)-\frac{RT_M\left(1-k\right)c_{\infty}}{v_0}p'\left(\psi\right)\left[\exp\left(u\right)-\frac{\Delta T}{mc_{\infty}}\right],
\end{equation}
\begin{equation}\label{eq-concen2}
	\frac{\partial c}{\partial t}=\nabla\cdot\left[D\bar{q}\left(\psi\right)c\nabla u-\bar{\mathbf{J}}_{at}-\bar{\mathbf{J}}\right].
\end{equation}
As discussed in some previous works \cite{Karma1998PRE,Karma2001PRL}, an additional freedom to obtain the desired thin-interface limit can be gained by replacing the function $p\left(\psi\right)$ in Eq. (\ref{eq-u}) and the function $\bar{p}\left(\psi\right)$ in temperature equation (\ref{eq-tem}) by another function $h\left(\psi\right)=\left(\psi-\psi_l\right)/\left(\psi_s-\psi_l\right)$, 
\begin{equation}
	u=\ln\frac{c/c_{\infty}}{1-\left(1-k\right)h\left(\psi\right)},\quad \frac{\partial}{\partial t}\left[\rho c_p\left(T-T_M\right)-\rho Lh\left(\psi\right)\right]=\nabla\cdot k_T\nabla T.
\end{equation}
What is more, the phase-field model [see Eq. (\ref{eq-concen2})] can be further completed by the following choices,
\begin{equation}
	\bar{q}\left(\psi\right)=\frac{\left(\psi_s-\psi\right)/\left(\psi_s-\psi_l\right)}{1-\left(1-k\right)h\left(\psi\right)},\quad \bar{\mathbf{J}}_{at}=-a\left(\psi\right)c_{\infty}\left(1-k\right)W_{\psi}\exp\left(u\right)\frac{\partial\psi}{\partial t}\frac{\nabla\psi}{|\nabla\psi|}.
\end{equation} 

Next, the following two dimensionless variables are introduced,
\begin{equation}
	U=\frac{\exp\left(u\right)-1}{1-k},\quad \theta=\frac{T-T_M-mc_{\infty}}{L_0/c_p^s},
\end{equation} 
where $L_0=L\left(T_M\right)$, $c_p^s$ is the specific heat per unit volume of solid phase. Then the phase-field, solute and heat transfer equations for solidification can be written as
\begin{subequations}\label{eq-PFE}
	\begin{equation}\label{eq-phe1}
		\tau\frac{\partial\psi}{\partial t}=W_{\psi}^2\nabla^2\psi-f'_2\left(\psi,T_M\right)-\partial_{\psi}f_3\left(\phi,\psi\right)-\lambda p'\left(\psi\right)\left(\theta+Mc_{\infty}U\right),
	\end{equation}
	\begin{equation}\label{eq-concen1}
		\frac{\partial C}{\partial t}=\left(1-k\right)\nabla\cdot\left[Dq\left(\psi\right)\nabla U-\mathbf{J}_{at}-\mathbf{J}\right],
	\end{equation}
	\begin{equation}\label{eq-tem1}
		\frac{\partial}{\partial t}\left[\rho\left(c_p\theta-\frac{L}{L_0}c_p^sh\left(\psi\right)\right)\right]=\nabla\cdot k_{T}\nabla\theta.
	\end{equation}
\end{subequations}
Here $M=-m\left(1-k\right)c_p^s/L_0$ is the scaled magnitude of the liquidus slope, $C=c/c_{\infty}$ is the dimensionless concentration, 
\begin{equation}
	q\left(\psi\right)=\frac{\psi_S-\psi}{\psi_s-\psi_l},\quad \mathbf{J}_{at}=-a\left(\psi\right)W_{\psi}\left[1+\left(1-k\right)U\right]\frac{\partial\psi}{\partial t}\frac{\nabla\psi}{|\nabla\psi|},\quad \mathbf{J}=Dq\left(\psi\right)C\frac{\sqrt{2\beta_{\phi}}\left(\phi_l-\phi\right)}{W_{\phi}}\frac{\nabla\phi}{|\nabla\phi|},
\end{equation}
the coupling constant $\lambda$ and the interpolation function $p\left(\psi\right)$ are defined by
\begin{equation}
	\lambda=-\frac{RT_M\left(1-k\right)L_0}{v_0mc_p^s}=\frac{L_0^2}{T_Mc_p^s},\quad
	p\left(\psi\right)=\frac{1}{\psi_s-\psi_l}\left[\frac{30}{\left(\psi_s-\psi_l\right)^4}\int\left(\psi_s-\psi\right)^2\left(\psi-\psi_l\right)^2\mathrm{d}\psi-\psi_l\right].
\end{equation}

In the commonly used phase-field models for solidification, the properties of solid dendrites and liquid melt, such as $\rho$, $c_p$ and $k_T$, are usually assumed to be matched, and $L\left(T\right)=L_0$ is chosen for simplicity. In this case, the temperature equation (\ref{eq-tem1}) can reduce to the following form,
\begin{equation}\label{eq-tem2}
	\frac{\partial\theta}{\partial t}=\nabla\cdot\alpha\nabla\theta+\frac{\partial h\left(\psi\right)}{\partial t},
\end{equation}
where $\alpha=k_T/\rho c_p$ is the thermal diffusivity. 
We also note that to keep consistent with the free boundary problem of alloy solidification, the parameters $\lambda$, $a\left(\psi\right)$ and $\tau$ can be determined from the asymptotic analysis of the thin-interface limit (see details in \ref{sec-analysis}). When taking $\beta_{\psi}=1/4$, $\psi_s=1$ and $\psi_l=-1$, one can obtain
\begin{equation}
	\lambda p'\left(\psi\right)=\frac{15L_0^2}{16T_Mc_p^s}\left(1-\psi^2\right)^2, \ u=\ln\frac{2C}{1+k-\left(1-k\right)\psi},\ \frac{\partial h\left(\psi\right)}{\partial t}=\frac{1}{2}\frac{\partial\psi}{\partial t},\ q\left(\psi\right)=\frac{1-\psi}{2},\ a\left(\psi\right)=\frac{1}{2\sqrt{2}}.
\end{equation}
With above expressions, the phase-field model [Eqs. (\ref{eq-phe1}), (\ref{eq-concen1}) and (\ref{eq-tem2})] is identical to some available ones \cite{Karma1998PRE,Karma2001PRL,Echebarria2004PRE,Ramirez2004PRE}.

To include the interfacial anisotropy, we set $W_{\psi}=W_0a_s\left(\mathbf{n}_{\psi}\right)$ with $\mathbf{n}_{\psi}=-\nabla\psi/|\nabla\psi|$ being the unit normal vector of liquid-solid interface, $a_s\left(\mathbf{n}_{\psi}\right)$ is the anisotropy function. In this case, the governing equations for binary alloy solidification in presence of gas bubbles can be expressed as
\begin{subequations}
	\begin{equation}\label{eq-ACE}
		\frac{\partial\phi}{\partial t}+\nabla\cdot\left(\phi\mathbf{v}\right)=\nabla\cdot M_{\phi}\left[\nabla\phi-\frac{\sqrt{2\beta_{\phi}}}{W_{\phi}}\left(\phi_l-\phi\right)\left(\phi-\phi_g\right)\frac{\nabla\phi}{|\nabla\phi|}\right],
	\end{equation}
	\begin{equation}\label{eq-AnACE}
		\tau_0a_s^2\left(\mathbf{n}_{\psi}\right)F\left(U\right)\frac{\partial\psi}{\partial t}=W_0^2\nabla\cdot\left[a_s^2\left(\mathbf{n}_{\psi}\right)\nabla\psi+\mathbf{N}\right]-f'_2\left(\psi,T_M\right)-\partial_{\psi}f_3\left(\phi,\psi\right)-\lambda p'\left(\psi\right)\left(\theta+Mc_{\infty}U\right),
	\end{equation}
	\begin{equation}
		\frac{\partial C}{\partial t}+\nabla\cdot\left(C\mathbf{v}\right)=\left(1-k\right)\nabla\cdot\left[Dq\left(\psi\right)\nabla U-\mathbf{J}_{at}-\mathbf{J}\right],
	\end{equation}
	\begin{equation}
		\frac{\partial}{\partial t}\left[\rho\left(c_p\theta-c_p^sh\left(\psi\right)\right)\right]+\nabla\cdot\left[\rho\left(c_p\theta-c_p^sh\left(\psi\right)\right)\mathbf{v}\right]=\nabla\cdot k_{T}\nabla\theta,
	\end{equation}
\end{subequations}  
where $\mathbf{v}$ is the velocity, $\tau_0$ is the relaxation time. We note that in this phase-field model, the convection effect is considered, and the complex fourth-order Cahn-Hilliard equation for gas-liquid interface capturing is replaced by the simple second-order conservative Allen-Cahn equation \cite{Chiu2011JCP}. The function $F\left(U\right)$ is determined through the thin-interface limit in \ref{sec-analysis}, and can be given by
\begin{equation}
	F\left(U\right)=\frac{1}{Le}+Mc_{\infty}\left[1+\left(1-k\right)U\right],
\end{equation} 
where $Le=\alpha/D$ is the Lewis number. $\mathbf{N}=\left(N_{\alpha}\right)$ is an anisotropic vector related to $a_s\left(\mathbf{n}_{\psi}\right)$,
\begin{equation}
	N_{\alpha}=|\nabla\psi|^2a_s\left(\mathbf{n}_{\psi}\right)\frac{\partial a_s\left(\mathbf{n}_{\psi}\right)}{\partial\left(\partial_{\alpha}\psi\right)},\quad \alpha=1,2,\cdots,d,
\end{equation} 
here the parameter $d$ represents the dimensionality.

\subsection{Incompressible Navier-Stokes equations}
To describe the melt flow, the following incompressible Navier-Stokes equations are considered,
\begin{subequations}
	\begin{equation}
		\nabla\cdot\mathbf{v}=0,
	\end{equation}
	\begin{equation}
		\frac{\partial\left(\rho\mathbf{v}\right)}{\partial t}+\nabla\cdot\left(\mathbf{mv}\right)=-\nabla P+\nabla\cdot\mu\left[\nabla\mathbf{v}+\left(\nabla\mathbf{v}\right)^\top\right]+\mathbf{F}_s+\rho\mathbf{f}+\mathbf{F}_b,
	\end{equation}
\end{subequations} 
where $\mathbf{m}$ is the mass flux, $P$ is the pressure, and $\mu$ is the dynamic viscosity. $\mathbf{F}_s$ is the surface tension force, $\mathbf{f}$ is the force caused by the fluid-solid interaction \cite{Zhan2023CiCP,Zhan2024PD}, and $\mathbf{F}_b$ represents the body force. The material property $\zeta$ in the entire computational domain can be given by
\begin{equation}
	\zeta=\frac{\psi-\psi_l}{\psi_s-\psi_l}\left(\zeta_s-\zeta_l\right)+\frac{\phi-\phi_g}{\phi_l-\phi_g}\left(\zeta_l-\zeta_g\right)+\zeta_g,
\end{equation}
where $\zeta_s$, $\zeta_l$ and $\zeta_g$ are the material properties (such as density, viscosity, specific heat per unit volume, and thermal conductivity) in pure solid, liquid and gas phases, respectively. However, in this work, $\zeta_s=\zeta_l$ is assumed to be consistent with the previous works for solidification process \cite{Wang1993PD,Karma1998PRE,Ramirez2004PRE}. Additionally, the mass flux $\mathbf{m}$ can be defined by $\mathbf{m}=\rho\mathbf{v}-\mathbf{D}\left(\phi\right)\left(\rho_l-\rho_g\right)/\left(\phi_l-\phi_g\right)$, in which the mass diffusion at gas-liquid interface is included \cite{Zhan2022PRE,Mirjalili2021JCP}, $\mathbf{D}\left(\phi\right)$ is the diffusion flux in phase-field equation (\ref{eq-ACE}), and is given by $M_{\phi}\left[\nabla\phi-\left(\sqrt{2\beta_{\phi}}/W_{\phi}\right)\left(\phi_l-\phi\right)\left(\phi-\phi_g\right)\nabla\phi/|\nabla\phi|\right]$.

Finally, to determine the form of the surface tension force $\mathbf{F}_s$, we use $b|\nabla\phi|^2$ to approximate the Dirac delta function $\delta$, 
\begin{equation}
	1=\int_{-\infty}^{+\infty}\delta\mathrm{d}x=\int_{-\infty}^{+\infty}b|\nabla\phi|^2\mathrm{d}x,\quad b=\frac{6W_{\phi}}{\left(\phi_l-\phi_g\right)^3\sqrt{2\beta_{\phi}}},
\end{equation}
where $|\nabla\phi|$ is calculated by the equilibrium profile of order parameter (\ref{eq-phiEq}). Then the surface tension force can be given by
\begin{equation}
	\mathbf{F}_s=-\sigma\delta\kappa_{\phi}\mathbf{n}_{\phi}=-b\sigma|\nabla\phi|^2\left(\nabla\cdot\mathbf{n}_{\phi}\right)\mathbf{n}_{\phi}=-b\sigma\left[\nabla^2\phi-\frac{\nabla\phi\cdot\nabla\left(|\nabla\phi|\right)}{|\nabla\phi|}\right]\nabla\phi=\frac{6\sigma}{\left(\phi_l-\phi_g\right)^3W_{\phi}\sqrt{2\beta_{\phi}}}\frac{\delta\mathcal{F}_{total}}{\delta\phi}\nabla\phi,
\end{equation}
here $\kappa_{\phi}$ and $\sigma$ are the curvature and the constant gas-liquid surface tension coefficient.

\section{Numerical method and validations}\label{sec-method}
In this section, we first introduce the numerical method, the mesoscopic LBM, and then conduct some numerical validations of the present phase-field model for gas-liquid two-phase flows and the thermosolutal dendritic growth. 

The unified evolution equation of LBM for convection-diffusion type equation and Navier-Stokes equations can be written as \cite{Zhan2023CiCP}
\begin{equation}
	f_i\left(\mathbf{x}+\mathbf{c}_i\Delta t,t+a_n\Delta t\right)=f_i\left(\mathbf{x},t\right)-\Lambda_{ij}\left(f_j-f_j^{eq}\right)\left(\mathbf{x},t\right)+\Delta t\left(F_i+\frac{\Delta t}{2}\hat{D}_iF_i\right)\left(\mathbf{x},t\right)+\Delta t\left(\delta_{ij}-\frac{\Lambda_{ij}}{2}\right)G_j\left(\mathbf{x},t\right),
\end{equation} 
where $f_i\left(\mathbf{x},t\right)$ is the distribution function of a macroscopic variable at position $\mathbf{x}$ and time $t$ along the $i$-th direction in the discrete velocity space, $f_i^{eq}\left(\mathbf{x},t\right)$ is the corresponding equilibrium distribution function. $F_i\left(\mathbf{x},t\right)$ and $G_i\left(\mathbf{x},t\right)$ are the distribution functions of source/force term and auxiliary term, respectively. $\bm{\Lambda}=\left(\Lambda_{ij}\right)$ is the collision matrix, $\hat{D}_i=a_n\partial_t+\bar{\gamma}\mathbf{c}_i\cdot\nabla$ with $\bar{\gamma}\in\{0,1\}$ being a parameter, $a_n$ is the relaxation of the time step $\Delta t$. Through choosing some specific moments of distribution functions, one can get the LBM for a specified macroscopic governing equation with the direct Taylor expansion (see Refs. \cite{Chai2020PRE,Zhan2022PRE,Zhan2023CiCP} for more details). Here the specific lattice Boltzmann models of LBM for the physical fields considered in this work are shown in \ref{sec-LBE}. 

In the following simulations, to perform some comparisons with the available works, the minima of the double-well profiles in Eq. (\ref{eq-douWell}) are set as $\phi=0,1$ and $\psi=\pm1$, i.e., $\phi_l=1$, $\phi_g=0$, $\psi_s=1$ and $\psi_l=-1$, and some other physical parameters are given by $\beta_{\phi}=1/4$, $\beta_{\psi}=1/4$, and $\beta_{inter}=3/4$. The half-way bounce-back scheme \cite{Ladd1994JFM1} is applied to treat the no-slip velocity and no-flux scalar boundary conditions, and the non-equilibrium extrapolation scheme \cite{Guo2002CP} is used to implement the Dirichlet velocity boundary conditions. For the problems with symmetric property, only the domain on one side of the symmetric axis is considered to save the computational resource, and the symmetric scheme for distribution functions is adopted \cite{Kruger2017}. 
 
\begin{figure}
	\centering
	\subfigure[$t=0$]{
		\begin{minipage}{0.18\linewidth}
			\centering
			\includegraphics[width=1.2in]{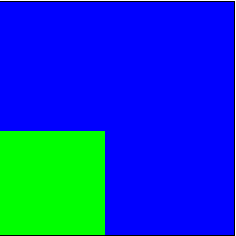}
	\end{minipage}}
	\subfigure[$t=80$]{
		\begin{minipage}{0.18\linewidth}
			\centering
			\includegraphics[width=1.2in]{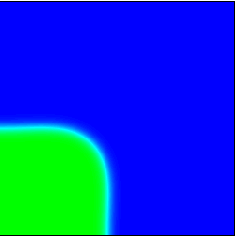}
	\end{minipage}}
	\subfigure[$t=240$]{
		\begin{minipage}{0.18\linewidth}
			\centering
			\includegraphics[width=1.2in]{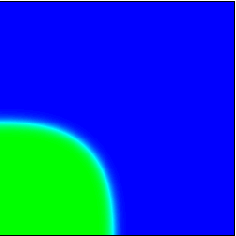}
	\end{minipage}}
	\subfigure[$t=400$]{
		\begin{minipage}{0.18\linewidth}
			\centering
			\includegraphics[width=1.2in]{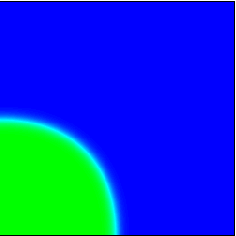}
	\end{minipage}}
	\subfigure[$t=810$]{
		\begin{minipage}{0.18\linewidth}
			\centering
			\includegraphics[width=1.2in]{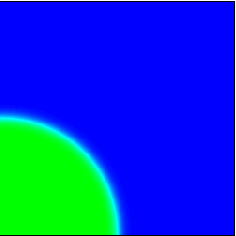}
	\end{minipage}}
	\caption{The deformation process of a square droplet.}
	\label{fig-droplet}
\end{figure}
 
\subsection{The deformation of a square droplet}
The deformation of a square droplet is a simple gas-liquid problem. With the increase of time, the square droplet would deform into a circle one under the action of the surface tension, and the pressure difference $\Delta P$ across the interface of the equilibrium droplet would obey the Young-Laplace law,
\begin{equation}\label{eq-laplace}
	\Delta P=\frac{\sigma}{R^{eq}},
\end{equation}
where $R^{eq}$ is the final radius of the droplet. Initially, a square liquid droplet with the length $D=\sqrt{\pi}R^{eq}$ is placed in the center of the physical domain $[-1,1]\times[-1,1]$, and the periodic boundary condition is applied on all boundaries. The material properties of the liquid droplet and the surrounding phase are given as $\rho_l=1000$, $\rho_g=1$, $\mu_l=10$, $\mu_g=0.01$, and $\sigma=0.01$. 

\begin{figure}
	\centering
	\includegraphics[width=2.2in]{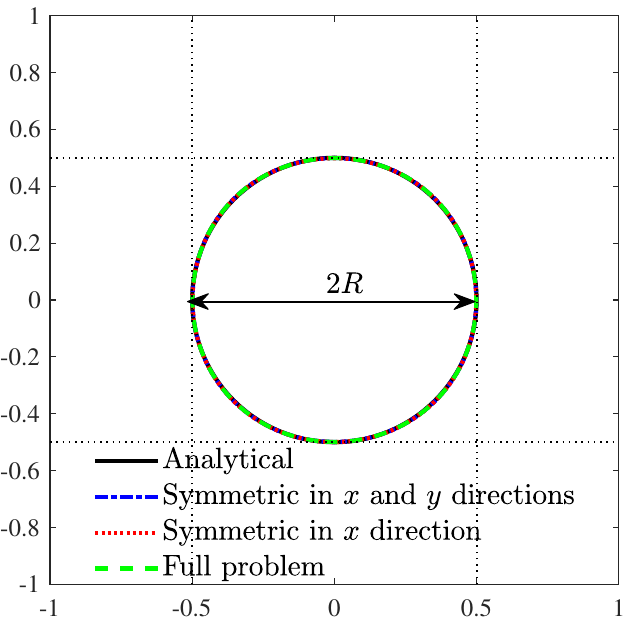}
	\caption{A comparison of numerical and analytical solutions in different cases.}
	\label{fig-dropCom}
\end{figure}

\begin{figure}
	\centering
	\includegraphics[width=3.0in]{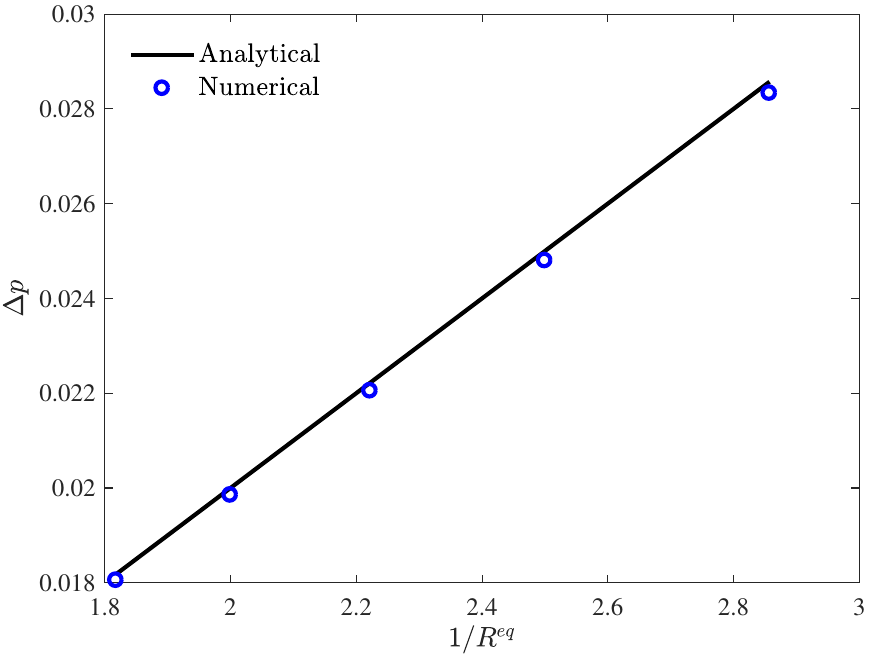}
	\caption{A comparison of numerical and analytical solutions of the pressure difference across the fluid interface of the equilibrium droplet.}
	\label{fig-laplace}
\end{figure}

To perform numerical simulations, the lattice spacing and time step are set to be $\Delta x=1/250$ and $\Delta t=\Delta x/50$, the other parameters are fixed as $M_{\phi}=0.1$ and $W_{\phi}=0.01$. Figure \ref{fig-droplet} shows the deformation process of the liquid droplet, and only the region of the first quadrant is simulated due to the symmetric property of this problem. We carry out a comparison of the results under different symmetric conditions and the full problem in Fig. \ref{fig-dropCom}. As shown in this figure, the numerical results of different cases are in agreement with the analytical solution $R^{eq}=0.5$, which confirms the accuracy of the symmetric scheme. To further examine the present method, we conduct some simulations under different radii, and plot the relation between the pressure difference and the droplet radius in Fig. \ref{fig-laplace}. From this figure, one can observe that there is a linear relationship between $\Delta P$ and $1/R^{eq}$, which is close to the analytical solution [Eq. (\ref{eq-laplace})]. 

\begin{figure}
	\centering
	\subfigure[$t=0$]{
		\begin{minipage}{0.18\linewidth}
			\centering
			\includegraphics[width=1.2in]{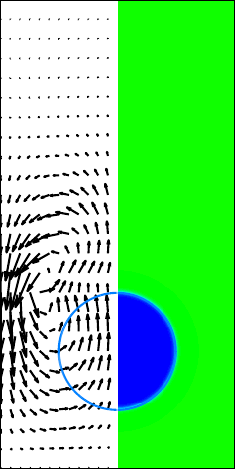}
		\end{minipage}}
	\subfigure[$t=2$]{
		\begin{minipage}{0.18\linewidth}
			\centering
			\includegraphics[width=1.2in]{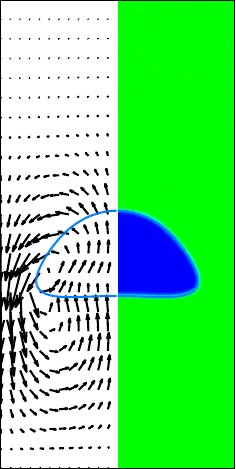}
		\end{minipage}}
	\subfigure[$t=4$]{
		\begin{minipage}{0.18\linewidth}
			\centering
			\includegraphics[width=1.2in]{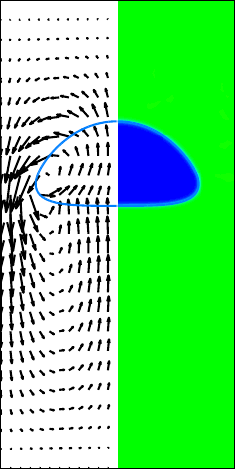}
		\end{minipage}}
	\subfigure[$t=6$]{
		\begin{minipage}{0.18\linewidth}
			\centering
			\includegraphics[width=1.2in]{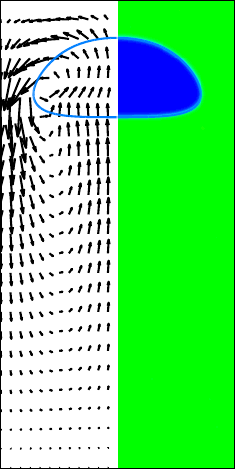}
		\end{minipage}}
	\subfigure[$t=8$]{
		\begin{minipage}{0.18\linewidth}
			\centering
			\includegraphics[width=1.2in]{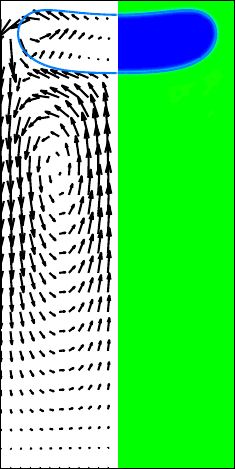}
		\end{minipage}}
	\caption{The evolution process of a single rising bubble.}
	\label{fig-bubble}
\end{figure}
\begin{figure}
	\centering
	\subfigure[]{
		\begin{minipage}{0.3\linewidth}
			\centering
			\includegraphics[width=1.8in]{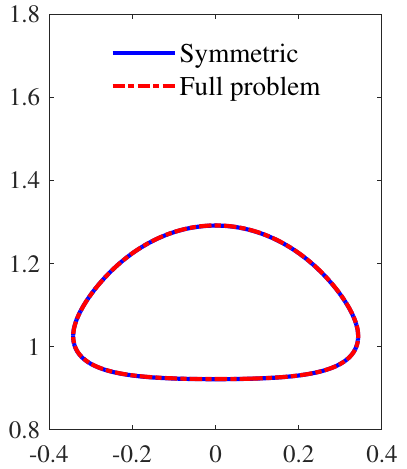}
	\end{minipage}}
	\subfigure[]{
		\begin{minipage}{0.3\linewidth}
			\centering
			\includegraphics[width=1.8in]{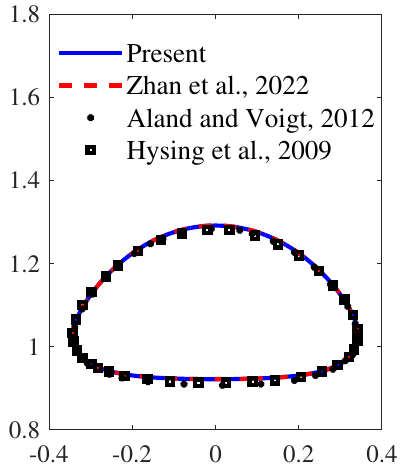}
	\end{minipage}}
	\caption{A comparison of the bubble shape at $t=3$.}
	\label{fig-bubbleComPhi}
\end{figure}
\begin{figure}
	\centering
	\subfigure[$t=0$]{
		\begin{minipage}{0.18\linewidth}
			\centering
			\includegraphics[width=1.2in]{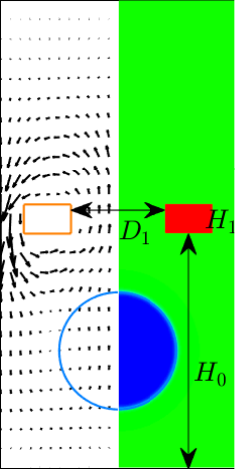}
	\end{minipage}}
	\subfigure[$t=2$]{
		\begin{minipage}{0.18\linewidth}
			\centering
			\includegraphics[width=1.2in]{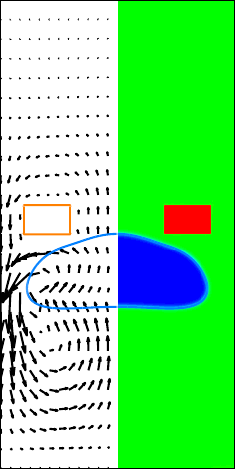}
	\end{minipage}}
	\subfigure[$t=4$]{
		\begin{minipage}{0.18\linewidth}
			\centering
			\includegraphics[width=1.2in]{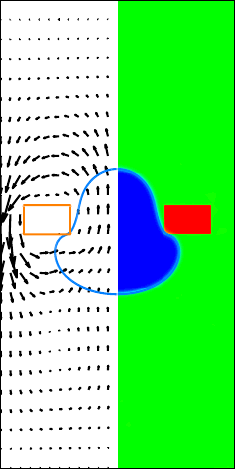}
	\end{minipage}}
	\subfigure[$t=6$]{
		\begin{minipage}{0.18\linewidth}
			\centering
			\includegraphics[width=1.2in]{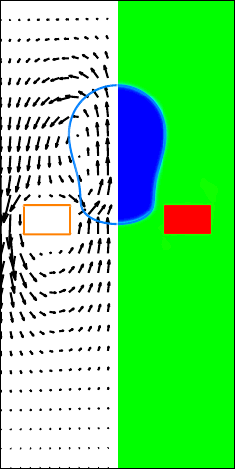}
	\end{minipage}}
	\subfigure[$t=8$]{
		\begin{minipage}{0.18\linewidth}
			\centering
			\includegraphics[width=1.2in]{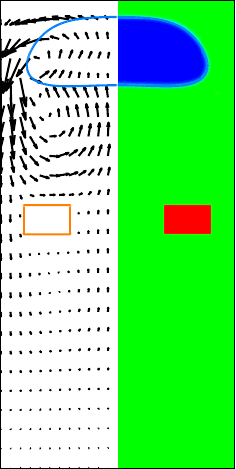}
	\end{minipage}}
	\caption{The evolution process of a single bubble rising through two obstacles.}
	\label{fig-bubbleObs}
\end{figure}
 
\subsection{A single rising bubble}
Now we consider the dynamics of a single rising bubble to further test the accuracy of the present method for gas-liquid flows. A circular bubble with the diameter $D$ is initially placed at position $\left(0,D\right)$ in a rectangular domain $[-D,D]\times[0,4D]$, and it would rise under the gravity $\mathbf{F}_b=\left(0,-\rho g\right)^\top$ with $g$ being the magnitude of the gravitational acceleration. For this problem, the periodic boundary condition is applied in $x$ direction, the no-flux boundary condition is imposed on the top and bottom boundaries, and the order parameter is initialized by
\begin{equation}
	\phi\left(x,y\right)=\frac{\phi_l+\phi_g}{2}+\frac{\phi_l-\phi_g}{2}\tanh\frac{\sqrt{x^2+\left(y-D\right)^2}-D/2}{\sqrt{2/\beta_{\phi}}W_{\phi}/\left(\phi_l-\phi_g\right)}.
\end{equation}    
To depict the dynamic behavior of the rising bubble, the following dimensionless Reynolds number and Bond number are used,
\begin{equation}
	Re=\frac{\rho_lD\sqrt{gD}}{\mu_l},\quad Bo=\frac{\rho_lgD^2}{\sigma}.
\end{equation}

\begin{figure}
	\centering
	\subfigure[]{
		\begin{minipage}{0.48\linewidth}
			\centering
			\includegraphics[width=3.0in]{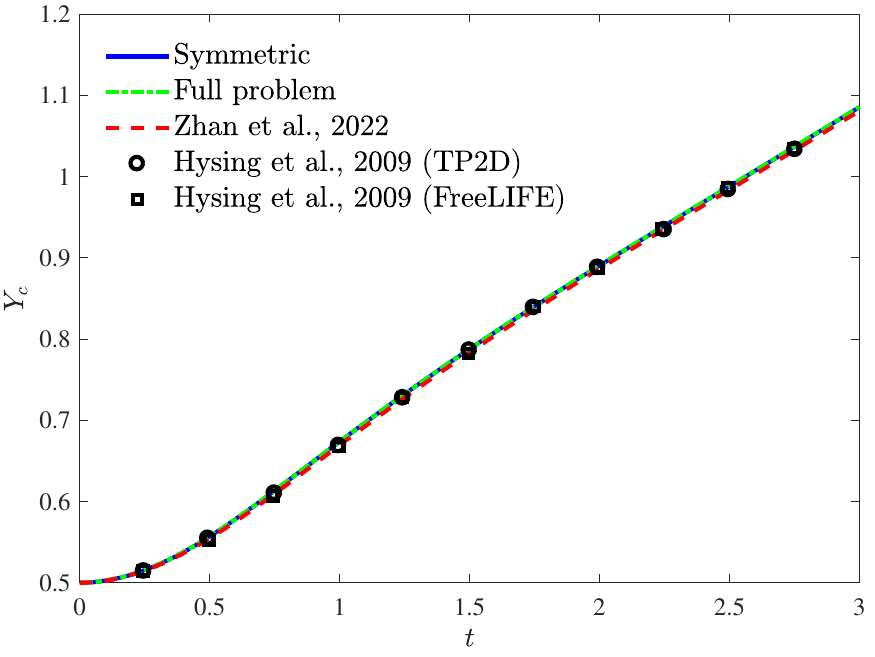}
	\end{minipage}}
	\subfigure[]{
		\begin{minipage}{0.48\linewidth}
			\centering
			\includegraphics[width=3.0in]{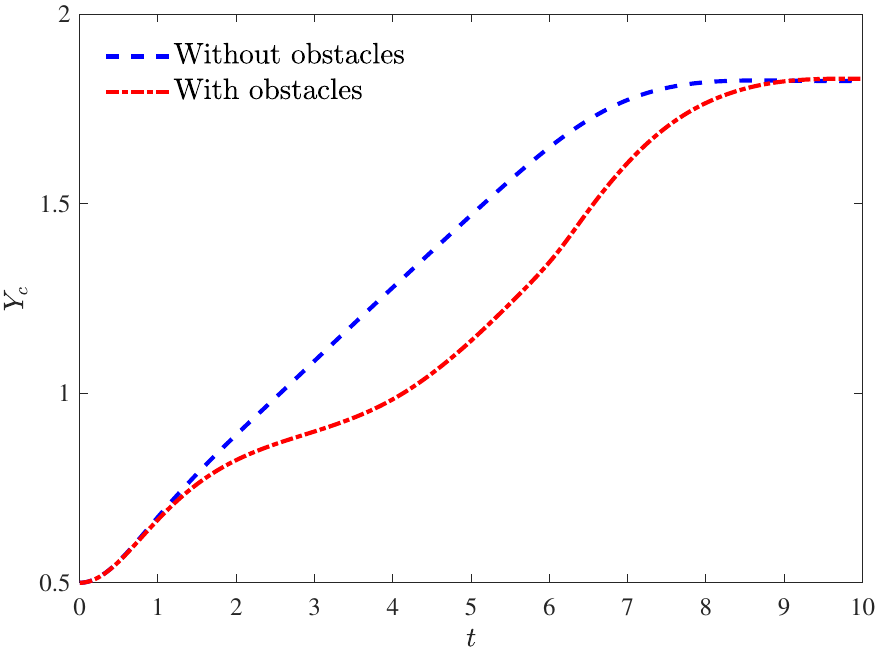}
	\end{minipage}}
	\caption{Comparisons of the mass center of the rising bubble without and with the effect of obstacles.}
	\label{fig-bubbleComYc}
\end{figure}

Here we consider the case of $Re=35$ and $Bo=10$, and the physical parameters are set to be $\rho_l=1000$, $\rho_g=100$, $\mu_l=10$, $\mu_g=1$, $g=0.98$ and $D=0.5$, which have been widely used in some previous works \cite{Aland2012IJNMF,Hysing2009IJNMF}. We perform some simulations in the region of the right side of the symmetric axis with $W_{\phi}=0.004$, $\Delta x=1/240$ and $\Delta t=\Delta x^2$, and show the dynamic behavior of the bubble in Fig. \ref{fig-bubble}. As seen from this figure, the bubble gradually rises and undergoes deformation under the action of gravity and surface tension. A further comparison of the bubble shape is plotted in Fig. \ref{fig-bubbleComPhi}, where the present results agree well with those reported in some available literature \cite{Zhan2022PRE,Aland2012IJNMF,Hysing2009IJNMF}. In addition, to give a quantitative comparison, we calculate the mass center of the bubble ($Y_c$) and show it as a function of time in Fig. \ref{fig-bubbleComYc}(a), in which a good agreement between the present results and some reported data \cite{Zhan2022PRE,Aland2012IJNMF,Hysing2009IJNMF} can be observed. 

In should be noted that in above simulations of gas-liquid flows, the computational domain is regular and the interaction between fluids and solid structure is not considered. To test the effect of interaction energy in Eq. (\ref{eq-f3Inter}), we place two square solid obstacles in the path of the rising bubble, and their influence can be included through introducing the order parameter $\psi$. The liquid-solid interface can be initialized by the form of Eq. (\ref{eq-psiEq}) with $W_{\psi}=W_{\phi}/4$. As shown in Fig. \ref{fig-bubbleObs}(a), the distance between the obstacle and the bottom of the domain is $H_0=2D$, the length of the orifice between two obstacles is $D_1=0.4D$, and the width of the obstacle is $H_1=0.25D$. Figure \ref{fig-bubbleObs} illustrates the dynamic evolution of the rising bubble in presence of the obstacles. From this figure, one can see that the rising bubble is hindered by the obstacles, and it deforms greatly to penetrate the orifice between two obstacles. After passing through the orifice, the bubble returns to a mushroom shape, like that of a classic rising bubble. Additionally, a comparison of the mass center of the bubble in Fig. \ref{fig-bubbleComYc}(b) also clearly illustrates the hindering effect of the obstacles.

\subsection{Thermosolutal dendritic growth}
In this part, we focus on the performance of the present phase-field model for the dendritic growth in the process of solidification. It is worth noting that when $Le=1$ and $Mc_{\infty}=0$, the model can be reduced to describe the thermal/iso-solutal dendritic growth, and the solutal/iso-thermal problems when $Le\rightarrow\infty$. To replace the condition $Le\rightarrow\infty$ in the numerical simulations, the temperature can be set as a given value $\theta_{sys}$ in the entire domain, and the temperature equation does not need to be solved. Through equaling the solutal and thermal capillary lengths, one can obtain \cite{Ramirez2004PRE}
\begin{equation}\label{eq-Tsys}
	\theta_{sys}=-Mc_{\infty}\frac{\Omega}{1-\left(1-k\right)\Omega},\quad Mc_{\infty}=1-\left(1-k\right)\Omega,
\end{equation}  
where the imposed solutal undercooling $\Omega$ is defined by
\begin{equation}
	\Omega=\frac{c_l^{eq}-c_{\infty}}{\left(1-k\right)c_l^{eq}}.
\end{equation}
If we further take $U=\Omega/\left[1-\left(1-k\right)\Omega\right]$ in the expression of relaxation time $\tau$, one can derive $F\left(U\right)=1$.

\begin{figure}
	\centering
	\subfigure[]{
		\begin{minipage}{0.35\linewidth}
			\centering
			\includegraphics[width=2.0in]{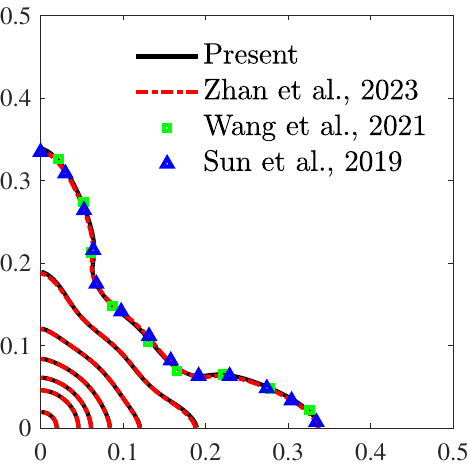}
	\end{minipage}}
	\subfigure[]{
		\begin{minipage}{0.35\linewidth}
			\centering
			\includegraphics[width=2.0in]{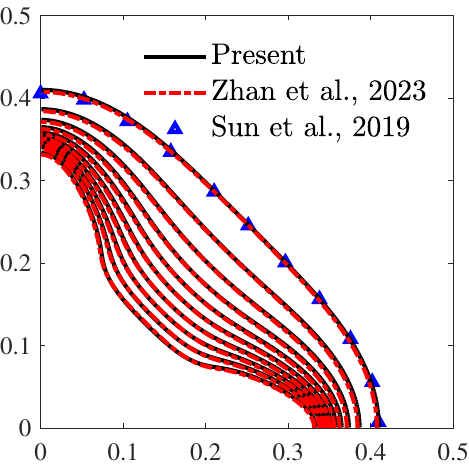}
	\end{minipage}}
	\caption{The interface profiles at $t/\tau_0=0,4,8,16,32,64,128$ (a) and isothermal lines from $\theta=-0.55$ to $\theta=-0.05$ with the increment of 0.05 at $t/\tau_0=128$ (b) in the thermal dendritic growth with pure diffusion.}
	\label{fig-thermal}
\end{figure}
\begin{figure}
	\centering
	\subfigure[]{
		\begin{minipage}{0.48\linewidth}
			\centering
			\includegraphics[width=3.0in]{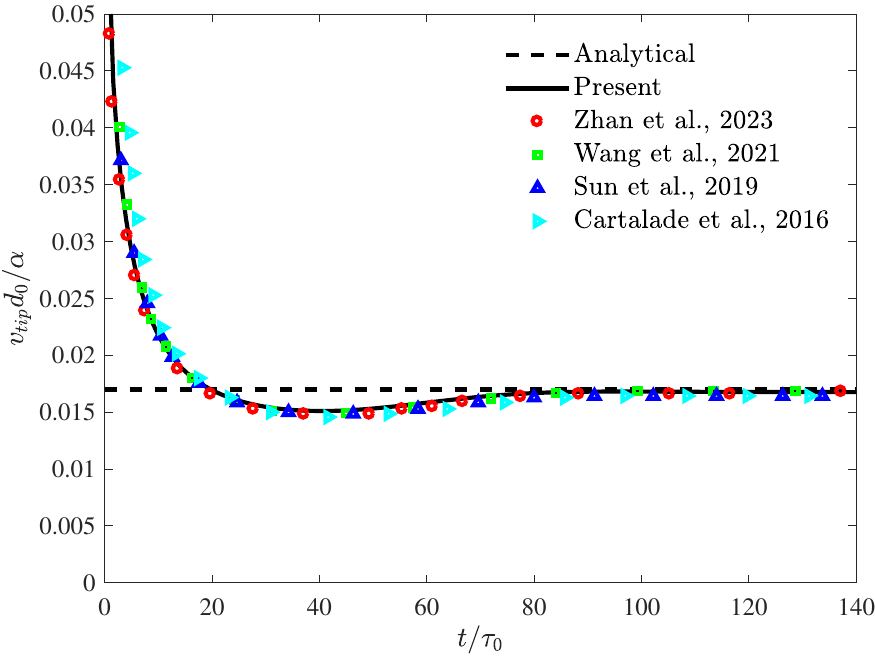}
	\end{minipage}}
	\subfigure[]{
		\begin{minipage}{0.48\linewidth}
			\centering
			\includegraphics[width=3.0in]{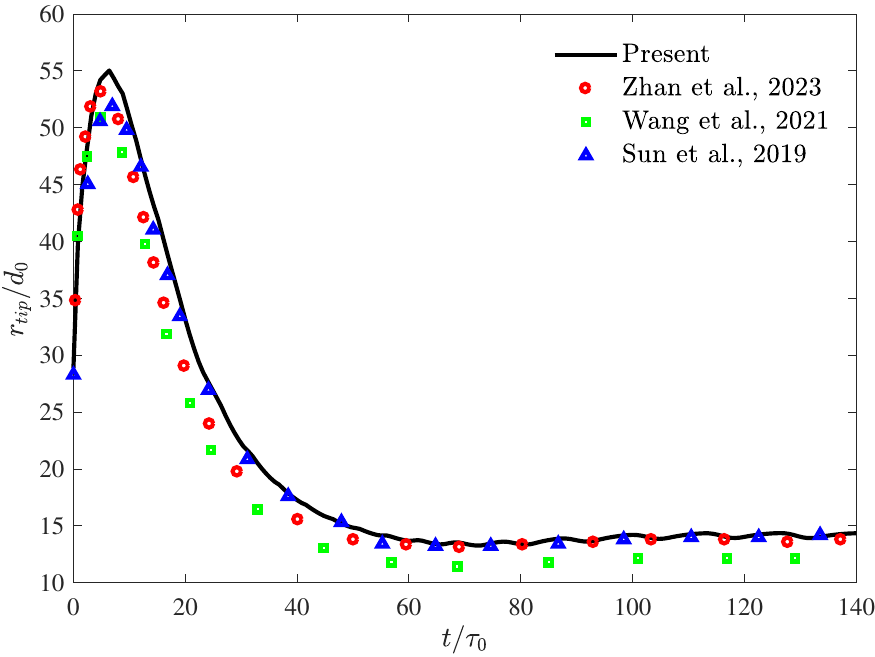}
	\end{minipage}}
	\caption{Evolutions of tip velocity (a) and tip radius (b) in the thermal dendritic growth with pure diffusion.}
	\label{fig-thermalTip}
\end{figure}

Additionally, for the cubic system of dendritic growth considered in this work, the anisotropy function of the interfacial energy $a_s\left(\mathbf{n}_{\psi}\right)$ is given by
\begin{equation}
	a_s\left(\mathbf{n}_{\psi}\right)=\left(1-3\epsilon_s\right)\left[1+\frac{4\epsilon_s}{1-3\epsilon_s}\sum_{\alpha=1}^{d}n_{\alpha}^4\right],
\end{equation}
where $n_{\alpha}$ is the $\alpha$-component of $\mathbf{n}_{\psi}$, and $\epsilon_s$ is the anisotropic strength.

\begin{figure}
	\centering
	\subfigure[]{
		\begin{minipage}{0.22\linewidth}
			\centering
			\includegraphics[width=1.4in]{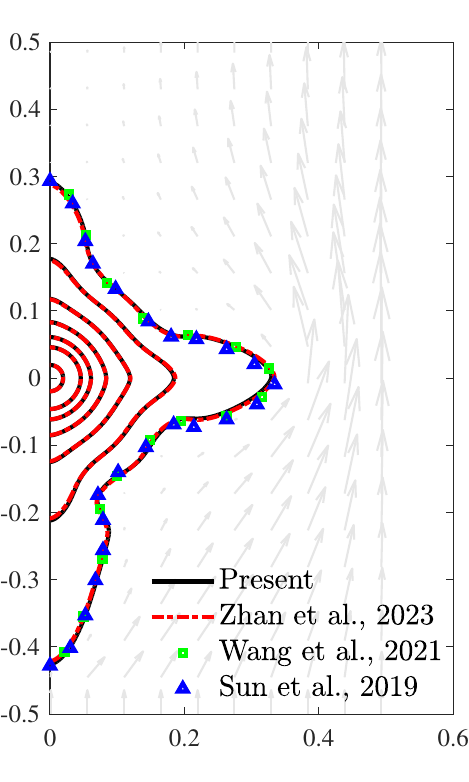}
	\end{minipage}}
	\subfigure[]{
		\begin{minipage}{0.22\linewidth}
			\centering
			\includegraphics[width=1.4in]{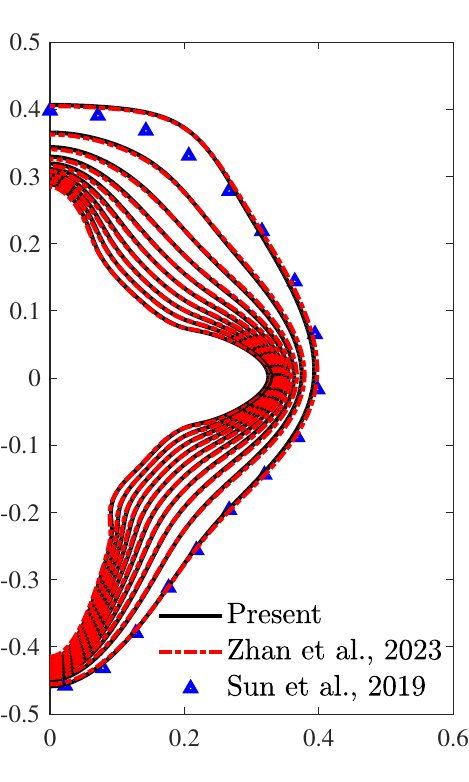}
	\end{minipage}}
	\subfigure[]{
		\begin{minipage}{0.5\linewidth}
			\centering
			\includegraphics[width=3.0in]{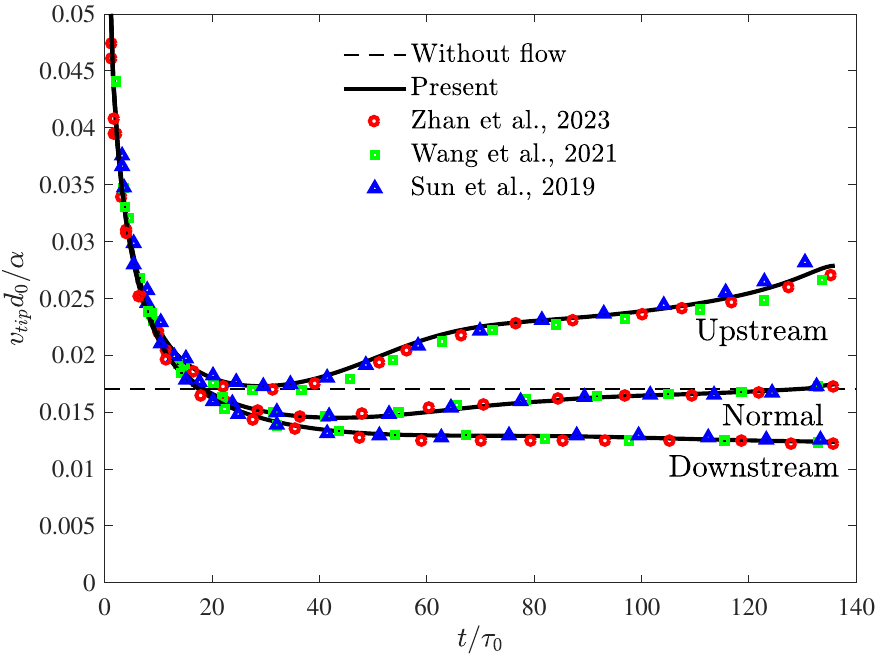}
	\end{minipage}}
	\caption{The interface profiles at $t/\tau_0=0,4,8,16,32,64,128$ superimposed with velocity field at $t/\tau_0=128$ (a), the isothermal lines from $\theta=-0.55$ to $\theta=-0.05$ with the increment of 0.05 at $t/\tau_0=128$ (b), and the evolution of the tip velocity (c) in the thermal dendritic growth with melt flow.}
	\label{fig-thermalU}
\end{figure}

In the case of the thermal dendritic growth of a pure substance, the solidification process is driven by the initial uniform undercooling $\theta=-0.55$, and the other physical parameters are set as $\epsilon_s=0.05$, the Peclet number $Pe_{D}=W_0^2/D\tau_0=LeW_0^2/\alpha\tau_0=0.25$, $W_0=1$ and $\tau_0=1$, then we can obtain $1/\lambda=a_2Pe_{D}$ and $d_0=a_1W_0/\lambda$. Initially, a circular seed with $R_s=10\Delta x$ is placed in the lower left corner of the square domain $\left[0,L\right]\times\left[0,L\right]$, the symmetric boundary condition is imposed on the bottom and left boundaries, while the no-flux boundary condition is applied on other boundaries. In the following simulations, the grid size is set to be $256\times256$, $\Delta x=0.4W_0$, $\Delta t=0.008\tau_0$. Figure \ref{fig-thermal} presents the evolution of the liquid-solid interface and the iso-thermal lines at $t/\tau_0=128$. From this figure, one can see that the seed grows into a dendrite with branches along the horizontal and vertical axes, and the gradients of temperature around the tips are much larger than those close to the dendritic root. To further quantitatively describe the dendritic growth, the tip velocity and tip radius are measured and plotted in Fig. \ref{fig-thermalTip}, where the present results agree well with the Green's function based analytical solution as well as some previous numerical data \cite{Sun2019IJHMT,Wang2021CMAME,Zhan2023CiCP,Cartalade2016CMA}. We also consider the case where the fluid is driven by the inlet velocity $\mathbf{v}_{in}=\left(0,W_0/\tau_0\right)$ in the domain $\left[0,L\right]\times\left[-L,L\right]$, and the viscosity of the melt can be determined by the Prandtl number $Pr=\mu_l/\rho_l\alpha=23.1$. As shown in Fig. \ref{fig-thermalU}, the dendritic arm in the upstream direction is longer and thicker than the normal and downstream arms, and the distribution of temperature field as well as the evolutions of tip velocities are consistent with those in Refs. \cite{Sun2019IJHMT,Wang2021CMAME,Zhan2023CiCP}.

\begin{figure}
	\centering
	\subfigure[]{
		\begin{minipage}{0.35\linewidth}
			\centering
			\includegraphics[width=2.0in]{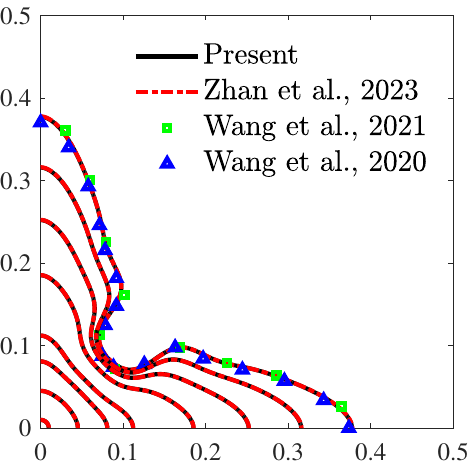}
	\end{minipage}}
	\subfigure[]{
		\begin{minipage}{0.35\linewidth}
			\centering
			\includegraphics[width=2.0in]{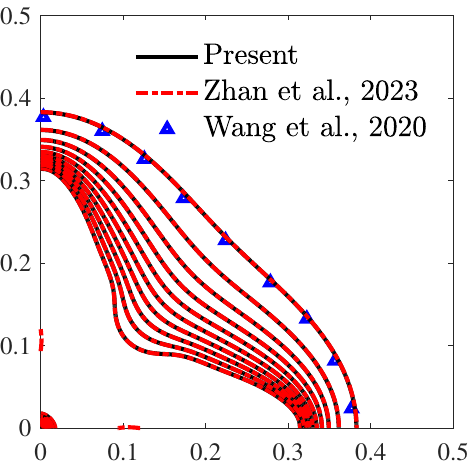}
	\end{minipage}}
	\caption{The interface profiles at $t/\tau_0=0,40,120,200,400,600,800,1000$ (a) and iso-solutal lines from $U=-0.55$ to $U=-0.05$ with the increment of 0.05 at $t/\tau_0=800$ (b) in the solutal dendritic growth with pure diffusion.}
	\label{fig-solutal}
\end{figure}

\begin{figure}
	\centering
	\subfigure[]{
		\begin{minipage}{0.48\linewidth}
			\centering
			\includegraphics[width=3.0in]{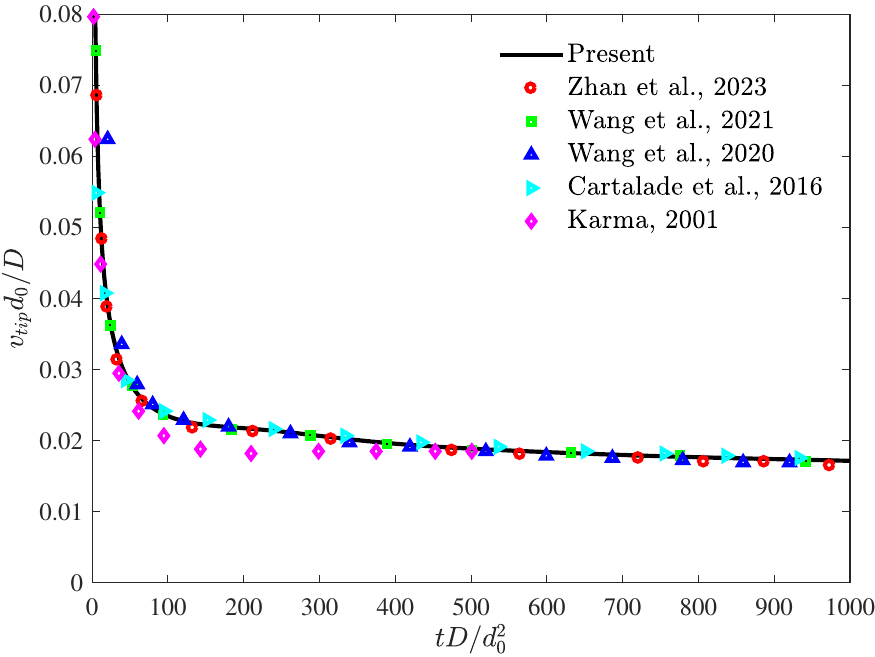}
	\end{minipage}}
	\subfigure[]{
		\begin{minipage}{0.48\linewidth}
			\centering
			\includegraphics[width=3.0in]{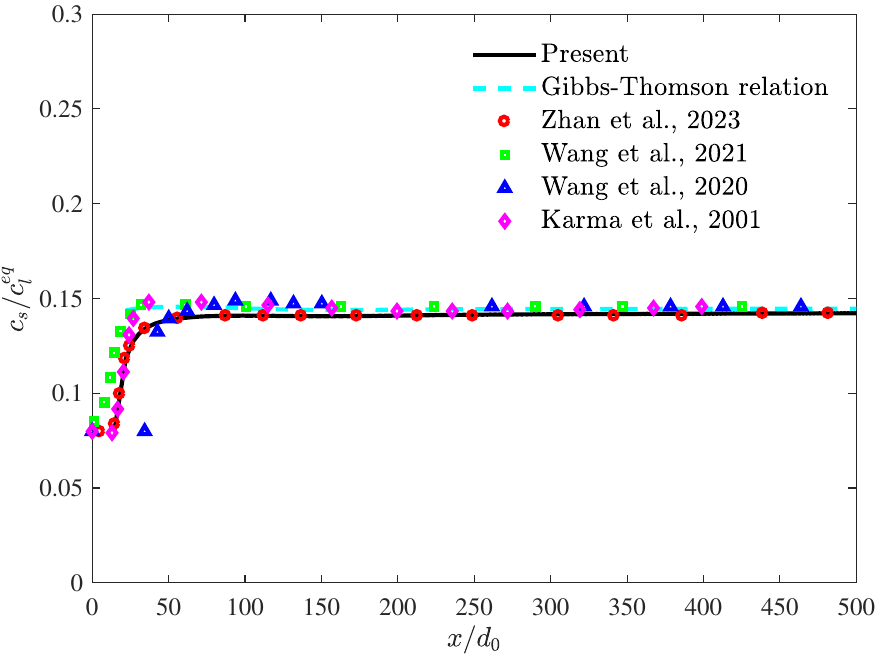}
	\end{minipage}}
	\caption{Evolution of tip velocity in the solutal dendritic growth with pure diffusion (a) and solute profiles on the solid side of interface along the central dendrite axis (b).}
	\label{fig-solutalTip}
\end{figure}

\begin{figure}
	\centering
	\includegraphics[width=2.0in]{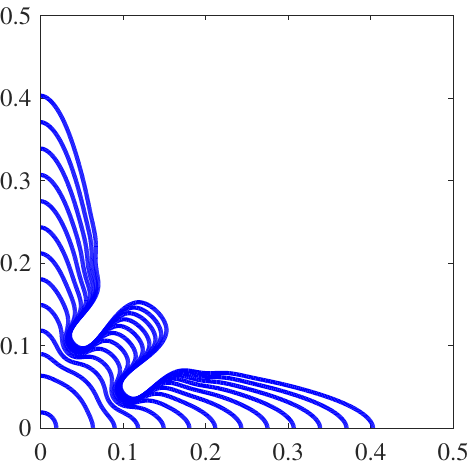}
	\caption{The interface profiles at every 20000 time steps in the solutal dendritic growth at the system temperature.}
	\label{fig-solutal-phi}
\end{figure}

\begin{figure}
	\centering
	\includegraphics[width=3.0in]{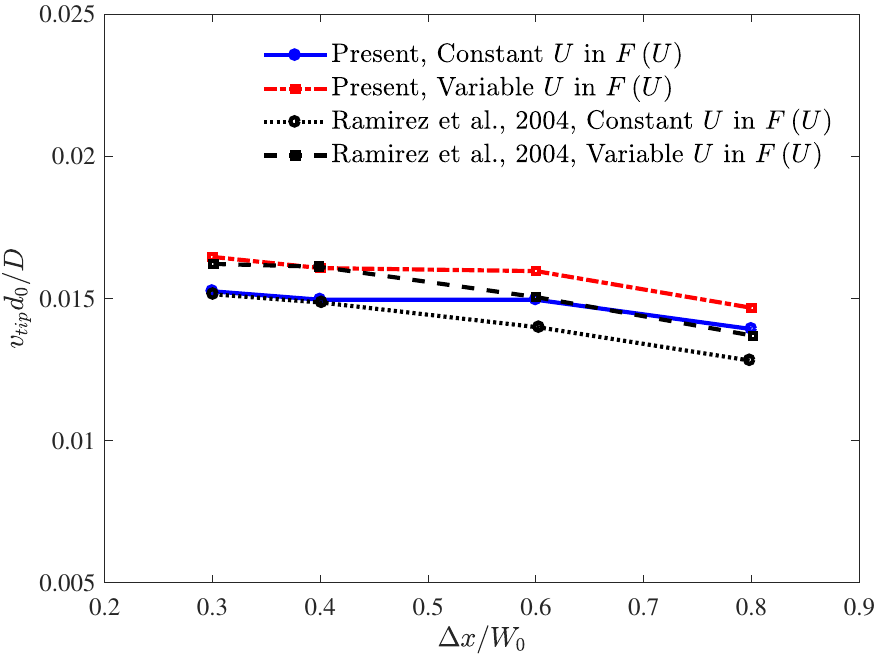}
	\caption{The tip velocity as a function of grid spacing.}
	\label{fig-solutal-dx}
\end{figure}

For the solutal dendritic growth of binary alloys under the iso-thermal condition, the computational domain, boundary conditions and the position of circular seed are set the same as those of the above iso-solutal case. Here we consider two cases: (A) the solutal dendritic growth without considering the effect of temperature and (B) solutal dendritic growth at the system temperature $\theta_{sys}$. These two cases correspond to the mathematical models in Ref. \cite{Karma2001PRL} and Ref. \cite{Ramirez2004PRE}, respectively. 
In the case (A), the physical parameters are set as $\Omega=0.55$, $k=0.15$, $\epsilon_s=0.02$, $D=2$, $\theta=0$ and $Mc_{\infty}=1$. We conduct a comparison of the evolutions of the interface morphology in Fig. \ref{fig-solutal}, where the initial radius of the seed is $R_s=10\Delta x$, the initial supersaturation is $U=-\Omega$, the grid size is $500\times500$, and time step is $\Delta t=0.02\tau_0$. From this figure, one can find that the present results are in agreement with the data in Ref. \cite{Zhan2023CiCP}. Figure \ref{fig-solutalTip} plots the tip velocity as a function of time and the dimensionless concentration on the solid side of the interface $c_s/c_l^{eq}$, which are also close to some available data \cite{Karma2001PRL,Cartalade2016CMA,Wang2020CMS,Wang2021CMAME,Zhan2023CiCP} and the Gibbs-Thomson relation $c_s/c_l^{eq}=k\left[1-\left(1-k\right)d_0/r_{tip}\right]$, where $r_{tip}$ changes with the displacement along the central dendrite axis. 
On the other hand, to give a comparison with the results in Ref. \cite{Ramirez2004PRE}, the system temperature $\theta_{sys}$ and $Mc_{\infty}$ are calculated by Eq. (\ref{eq-Tsys}). In case (B), the grid size and time step are set to be $800\times800$ and $\Delta t=0.008\tau_0$, the initial radius of the seed is fixed as $R_s=44d_0$ and the supersaturation is $U=0$ such that $C=1-\left(1-k\right)h\left(\psi\right)$. We perform some simulations, and plot the evolution of liquid-solid interface in Fig. \ref{fig-solutal-phi}, in which the morphology of the dendrite is close to the results in Ref. \cite{Ramirez2004PRE}. We further present a quantitative comparison of the tip velocity under different grid sizes ($\Delta x/W_0=0.3,0.4,0.6,0.8$) but same physical size of computational domain. As seen from Fig. \ref{fig-solutal-dx}, compared to Ref. \cite{Ramirez2004PRE}, the present results have a similar convergence in grid spacing.

\begin{figure}
	\centering
	\includegraphics[width=3.0in]{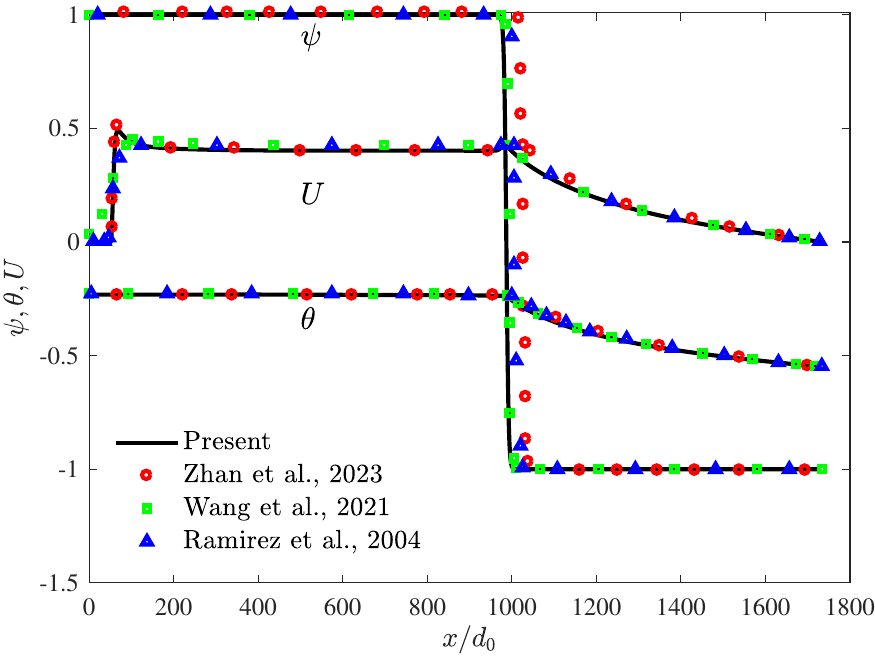}
	\caption{The profiles of $\phi$, $U$, $\theta$ along the central dendrite axis at $tD/d_0^2=470000$.}
	\label{fig-thermosolutal}
\end{figure}

\begin{figure}
	\centering
	\subfigure[]{
		\begin{minipage}{0.48\linewidth}
			\centering
			\includegraphics[width=3.0in]{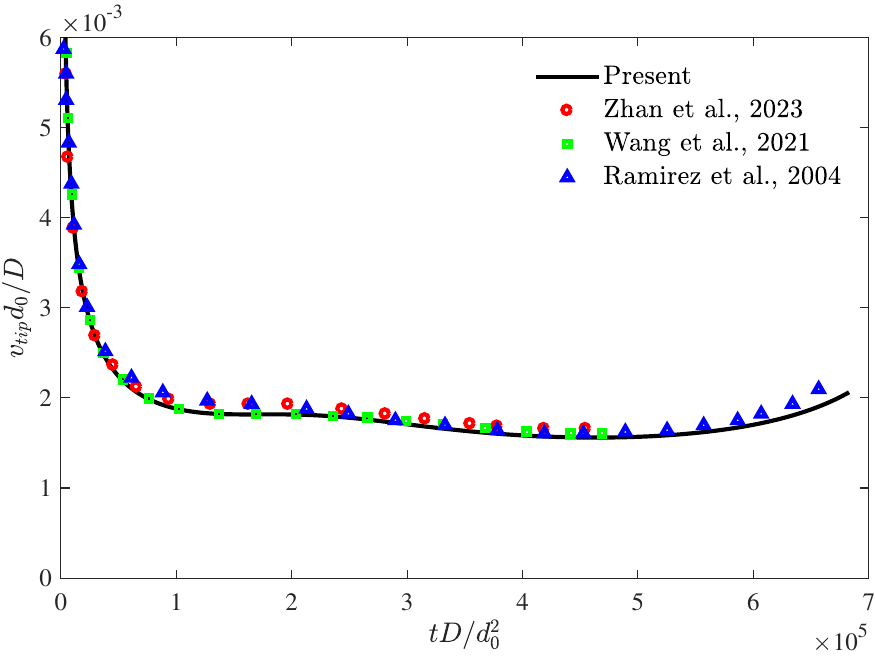}
	\end{minipage}}
	\subfigure[]{
		\begin{minipage}{0.48\linewidth}
			\centering
			\includegraphics[width=3.0in]{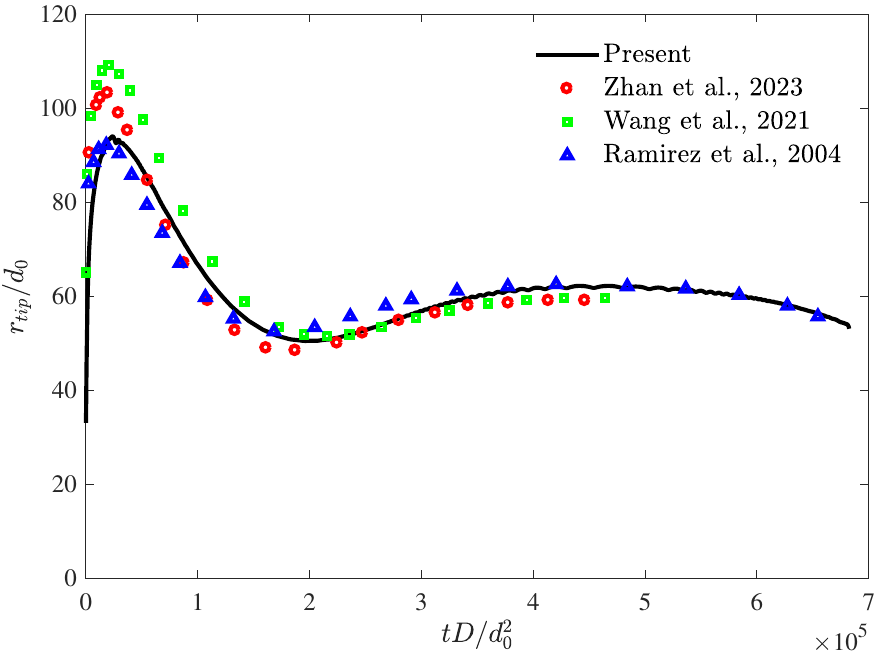}
	\end{minipage}}
	\caption{Evolutions of tip velocity (a) and tip radius (b) in the thermosolutal dendritic growth with pure diffusion.}
	\label{fig-thermosolutalTip}
\end{figure}

Finally, we consider the dendritic growth of a binary alloy in an undercooled melt. Compared to the iso-thermal case, here the temperature equation needs to be solved, and the far-field Dirichlet boundary conditions $\psi=\psi_l$, $C=1$ and $\theta=-\Omega$ are imposed to the top and right boundaries. In our simulations, the grid size is $1200\times1200$, the time step is $\Delta t=0.018\tau_0$, the initial seed radius is $R_s=65d_0$, and the Lewis number is $Le=1$. We give a comparison of the profiles of some variables along the central dendrite axis in Fig. \ref{fig-thermosolutal} where $tD/d_0^2=470000$. From this figure, one can observe that the present results are in good agreement with those in Refs. \cite{Karma2001PRL,Wang2021CMAME,Zhan2023CiCP}. In addition, the evolutions of the tip velocity and radius are shown in Fig. \ref{fig-thermosolutalTip}, and one can find that they are close to the data reported in Refs. \cite{Karma2001PRL,Wang2021CMAME,Zhan2023CiCP}. Here the early deviation in Fig. \ref{fig-thermosolutalTip}(b) may be caused by the evaluation method of the tip radius.

These above validations indicate that the present diffuse-interface model can be used to describe the gas-liquid flows in complex geometries and the solidification of binary alloy, respectively. In the following, it will be adopted to study more complex solidification process in presence of gas bubbles. 


\section{Results and discussion}\label{sec-results}
In this section, we will focus on the problem of thermosolutal dendritic growth in the presence of gas bubbles. In the computational domain $Lx\times Ly=200\times400$ with symmetric boundary condition in $x$-direction, a bubble with the diameter $D=40$ is initially located at $\left(x_0,y_0\right)$, and two seeds with the radius $R_s=10\Delta x$ are placed at the left-bottom and right-bottom corners. The bubble would rise under the gravity, and simultaneously, the seeds would grow in the undercooled melt with no-flux boundary condition applied on the bottom boundary and far-field Dirichlet boundary condition ($\psi=\psi_l$, $C=1$, and $\theta=-\Omega$) imposed on the top boundary. To depict this problem, we also use above-defined dimensionless numbers, and set them to be $Re=35$, $Bo=500$, $Le=50$, $Pe_{D}=1$, and $Pr=23.1$. In the following simulations, some physical parameters are fixed as $W_0=1$, $\tau_0=1$, $k=0.15$, $\epsilon_s=0.02$, $\rho_l=2000$, $\rho_g=1$, $\mu_l/\mu_g=100$, $g=1/D$, and the effects of undercooling and bubble position will be studied under $\Delta x=0.4W_0$ and $\Delta t=0.004\tau_0$. 

\begin{figure}
	\centering
	\subfigure[$t/\tau_0=2000$]{
		\begin{minipage}{0.24\linewidth}
			\centering
			\includegraphics[width=1.6in]{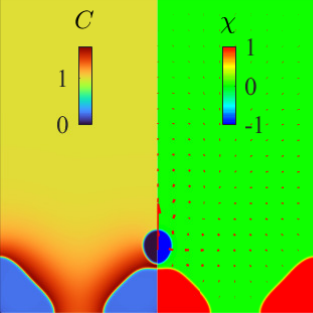}
	\end{minipage}}
	\subfigure[$t/\tau_0=4000$]{
		\begin{minipage}{0.24\linewidth}
			\centering
			\includegraphics[width=1.6in]{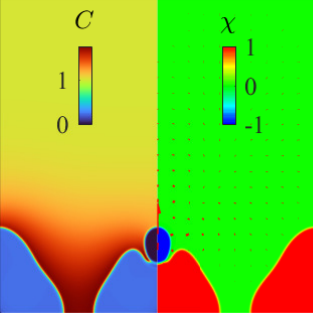}
	\end{minipage}}
	\subfigure[$t/\tau_0=6000$]{
		\begin{minipage}{0.24\linewidth}
			\centering
			\includegraphics[width=1.6in]{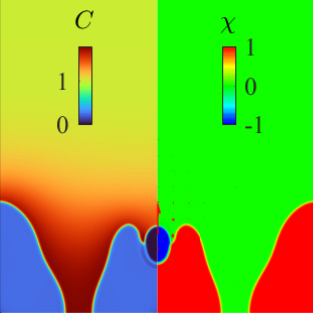}
	\end{minipage}}
	\subfigure[$t/\tau_0=10000$]{
		\begin{minipage}{0.24\linewidth}
			\centering
			\includegraphics[width=1.6in]{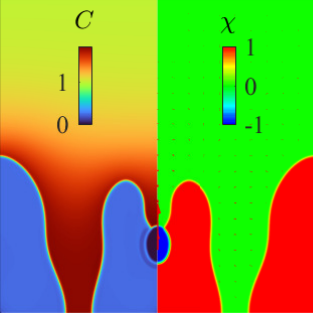}
	\end{minipage}}
	\caption{The evolution of thermosolutal dendritic growth in presence of a gas bubble, $\chi=\left(\psi-\psi_l\right)/\left(\psi_s-\psi_l\right)-\left(\phi_l-\phi\right)/\left(\phi_l-\phi_g\right)$.}
	\label{fig-fullcoupled}
\end{figure}

Before performing any discussion, we first show the evolutions of a basic case with $\Omega=0.55$ and $\left(x_0,y_0\right)=\left(0,Ly/5\right)$ in Fig. \ref{fig-fullcoupled}. As seen from this figure, the seeds grow into dendrites faster than the rate of gas bubble rising, such that the gas bubble is pinned by the central dendrite at about $t/\tau_0=4000$ and then trapped with time increases. When the gas bubble is not totally trapped yet, it would deform obviously under the combined action of gravity and solid dendrite.

\subsection{The effect of undercooling}
\begin{figure}
	\centering
	\subfigure[$\Omega=0.5$]{
		\begin{minipage}{0.24\linewidth}
			\centering
			\includegraphics[width=1.6in]{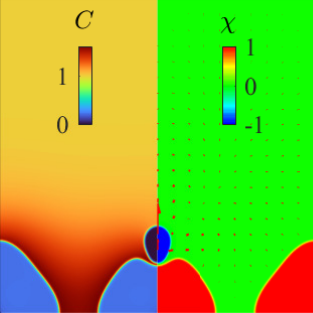}
	\end{minipage}}
	\subfigure[$\Omega=0.55$]{
		\begin{minipage}{0.24\linewidth}
			\centering
			\includegraphics[width=1.6in]{fig-n2-4000.pdf}
	\end{minipage}}
	\subfigure[$\Omega=0.65$]{
		\begin{minipage}{0.24\linewidth}
			\centering
			\includegraphics[width=1.6in]{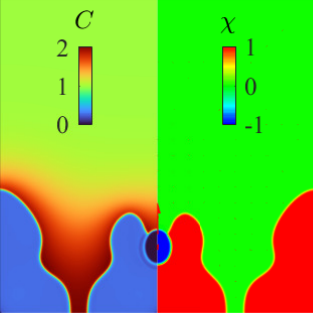}
	\end{minipage}}
	\caption{The snapshots of concentration and phase fields superimposed with velocity field under different values of undercooling $\Omega$.}
	\label{fig-undercool-4000}
\end{figure}

\begin{figure}
	\centering
	\subfigure[]{
		\begin{minipage}{0.48\linewidth}
			\centering
			\includegraphics[width=3.0in]{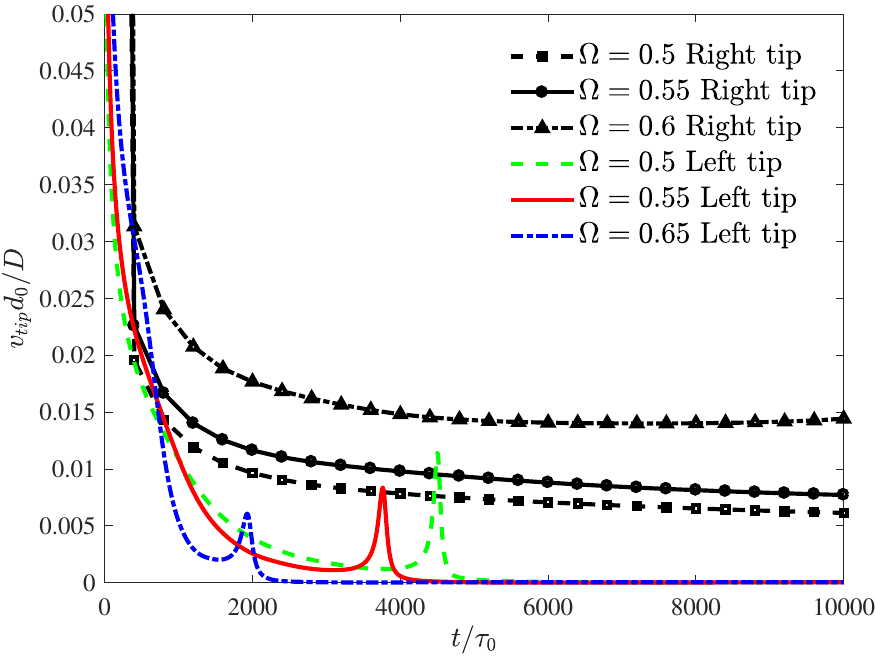}
	\end{minipage}}
	\subfigure[]{
		\begin{minipage}{0.48\linewidth}
			\centering
			\includegraphics[width=3.0in]{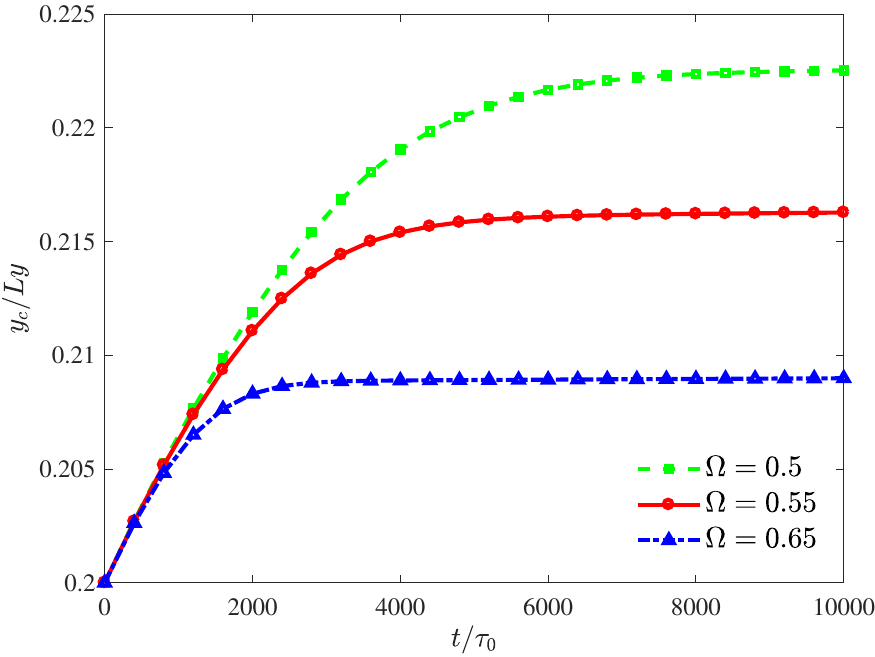}
	\end{minipage}}
	\caption{Evolutions of tip velocities (a) and the mass center of bubble (b) in the thermosolutal dendritic growth.}
	\label{fig-undercool-Com}
\end{figure}

We first consider three values of undercooling, i.e., $\Omega=0.5$, $0.55$ and $0.65$, and present the snapshots of concentration and phase fields in Fig. \ref{fig-undercool-4000} where $t/\tau_0=4000$. As shown in this figure, a larger value of undercooling can cause a faster solidification process, and the gas bubble is trapped earlier. We also plot the evolutions of the tip velocities of the left dendrite ($x=0$) and the right dendrite ($x=Lx$) in Fig. \ref{fig-undercool-Com}(a). In this figure, the tip velocities at $x=Lx$ decrease with the increase of time and would converge to the steady values, while those at $x=0$ exhibit dramatic changes. Due to the blocking effect of the bubble in the solidification direction, the tip velocities decrease faster before the dendrites kiss the bubble. After that, the values of $v_{tip}$ have a sudden increase and then decrease rapidly to zero under the dominant effect of the interaction between solid dendrite and gas bubble. Furthermore, the time sequence at which the change occurs also reflects the effect of the undercooling. Additionally, the dynamics of gas bubble can be quantitatively described by the mass center of the bubble ($y_c/Ly$) versus time in Fig. \ref{fig-undercool-Com}(b). It is clear that the mass center of bubble first increases under the action of gravity, and then reaches to a steady value after the bubble trapped by the dendrite. In addition, one can also see from this figure that a lower value of the undercooling also induces a higher position of the bubble because of the slower pinning time.

 
\subsection{The effect of initial bubble position}
\begin{figure}
	\centering
	\subfigure[$x_0=Lx/8$]{
		\begin{minipage}{0.24\linewidth}
			\centering
			\includegraphics[width=1.6in]{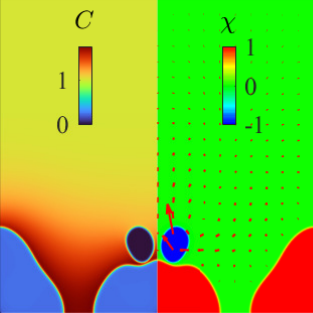}
	\end{minipage}}
	\subfigure[$x_0=Lx/4$]{
		\begin{minipage}{0.24\linewidth}
			\centering
			\includegraphics[width=1.6in]{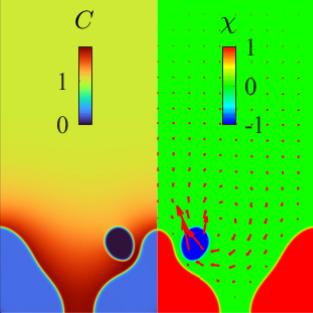}
	\end{minipage}}
	\subfigure[$x_0=3Lx/8$]{
		\begin{minipage}{0.24\linewidth}
			\centering
			\includegraphics[width=1.6in]{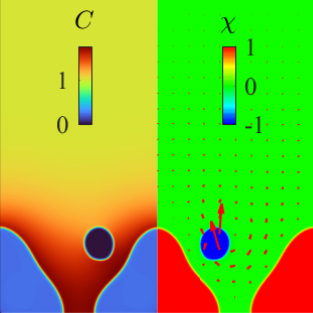}
	\end{minipage}}
	\subfigure[$x_0=Lx/2$]{
		\begin{minipage}{0.24\linewidth}
			\centering
			\includegraphics[width=1.6in]{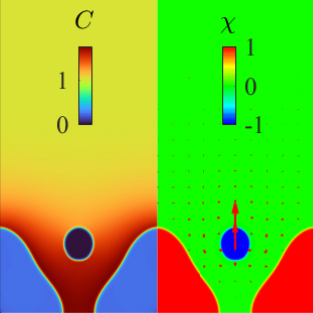}
	\end{minipage}}

	\subfigure[$x_0=Lx/8$]{
		\begin{minipage}{0.24\linewidth}
			\centering
			\includegraphics[width=1.6in]{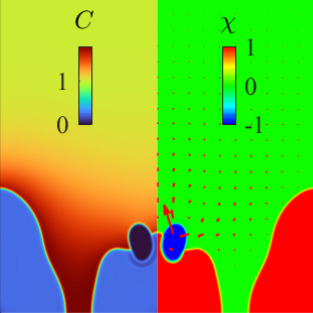}
	\end{minipage}}
	\subfigure[$x_0=Lx/4$]{
		\begin{minipage}{0.24\linewidth}
			\centering
			\includegraphics[width=1.6in]{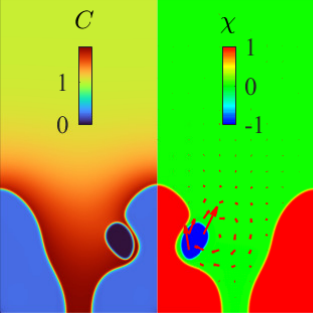}
	\end{minipage}}
	\subfigure[$x_0=3Lx/8$]{
		\begin{minipage}{0.24\linewidth}
			\centering
			\includegraphics[width=1.6in]{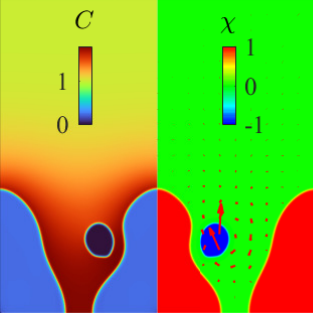}
	\end{minipage}}
	\subfigure[$x_0=Lx/2$]{
		\begin{minipage}{0.24\linewidth}
			\centering
			\includegraphics[width=1.6in]{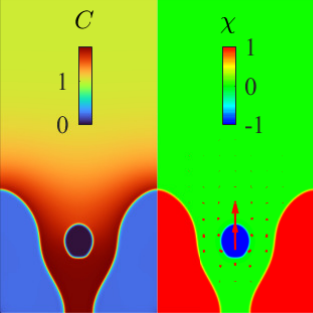}
	\end{minipage}}

	\subfigure[$x_0=Lx/8$]{
		\begin{minipage}{0.24\linewidth}
			\centering
			\includegraphics[width=1.6in]{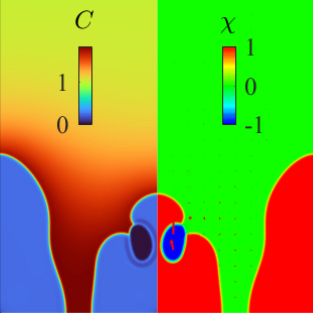}
	\end{minipage}}
	\subfigure[$x_0=Lx/4$]{
		\begin{minipage}{0.24\linewidth}
			\centering
			\includegraphics[width=1.6in]{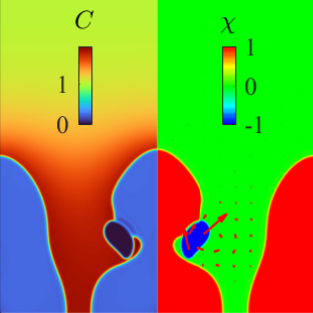}
	\end{minipage}}
	\subfigure[$x_0=3Lx/8$]{
		\begin{minipage}{0.24\linewidth}
			\centering
			\includegraphics[width=1.6in]{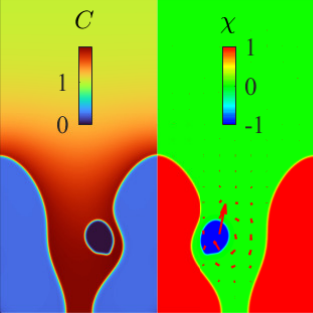}
	\end{minipage}}
	\subfigure[$x_0=Lx/2$]{
		\begin{minipage}{0.24\linewidth}
			\centering
			\includegraphics[width=1.6in]{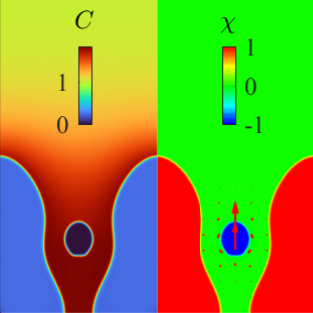}
	\end{minipage}}
	\caption{The snapshots of concentration and phase fields superimposed with velocity field with different initial positions of the gas bubble [ $t/\tau_0=4000$ (top row), $t/\tau_0=7000$ (middle row) and $t/\tau_0=10000$ (bottom row)].}
	\label{fig-position-times}
\end{figure}

\begin{figure}
	\centering
	\subfigure[]{
		\begin{minipage}{0.48\linewidth}
			\centering
			\includegraphics[width=3.0in]{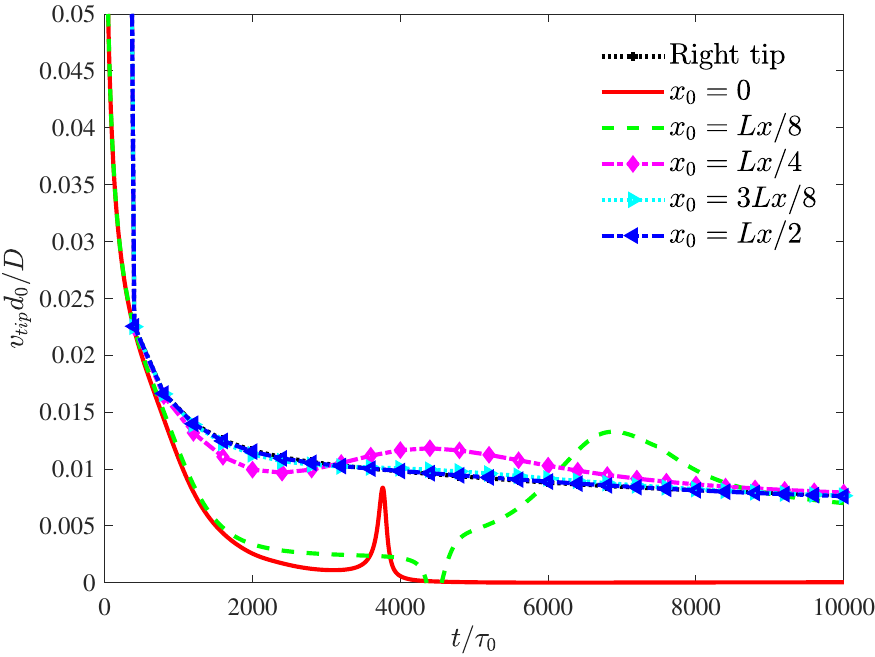}
	\end{minipage}}
	\caption{Evolutions of tip velocities in the thermosolutal dendritic growth with gas bubbles.}
\label{fig-position-ComV}
\end{figure}

\begin{figure}
	\centering
	\subfigure[]{
		\begin{minipage}{0.48\linewidth}
			\centering
			\includegraphics[width=3.0in]{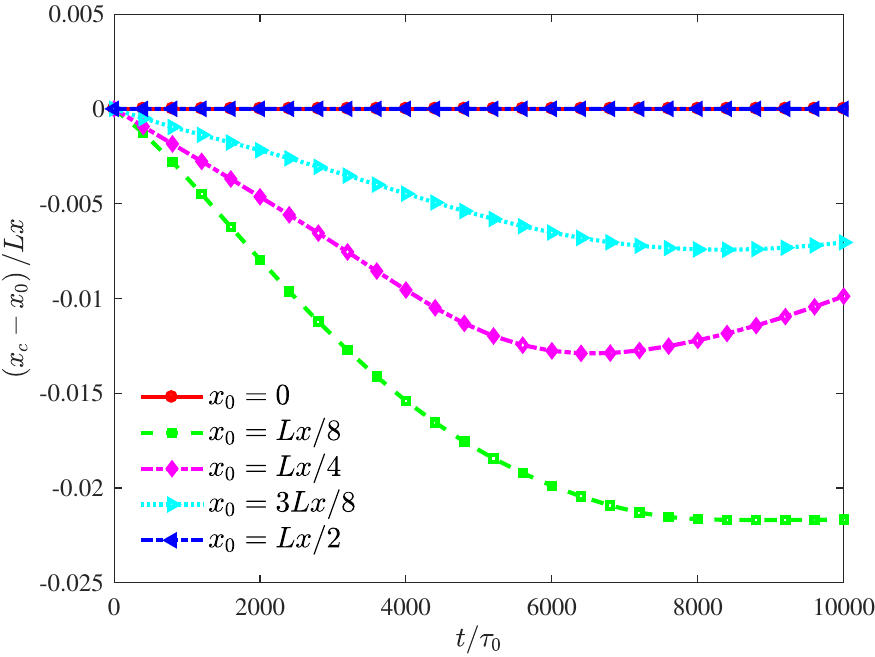}
	\end{minipage}}
	\subfigure[]{
		\begin{minipage}{0.48\linewidth}
			\centering
			\includegraphics[width=3.0in]{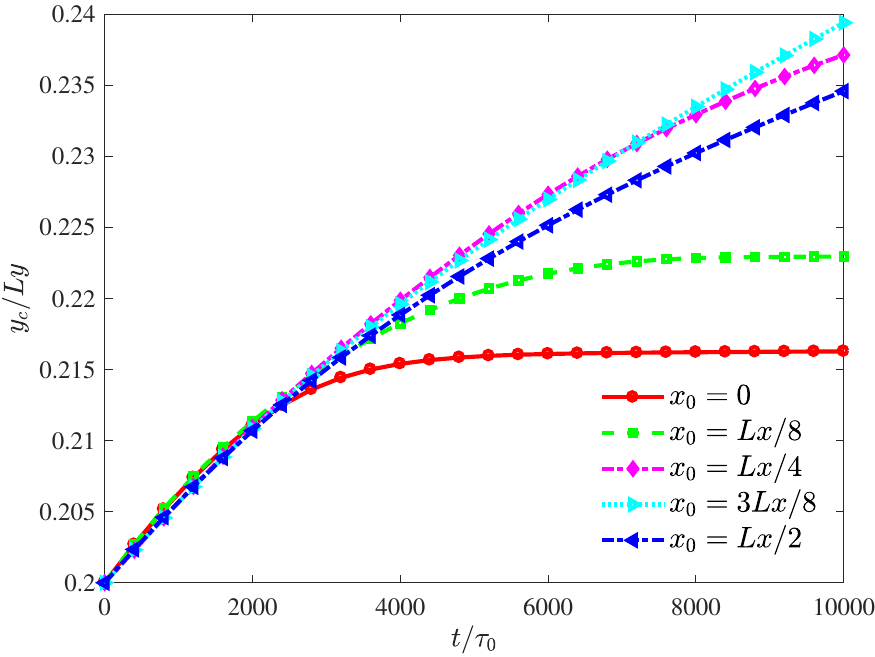}
	\end{minipage}}
	\caption{Evolutions of mass center of bubble in the thermosolutal dendritic growth [(a) the normalized horizontal component, (b) the normalized vertical component].}
	\label{fig-position-ComP}
\end{figure}

Now we investigate the influence of initial bubble position on the growth of the dendrites and the interaction between dendrites and bubble. The bubble is initially placed at $y_0=Ly/5$ with different horizontal positions ($x_0=0$, $Lx/8$, $Lx/4$, $3Lx/8$ and $Lx/2$). As shown in Fig. \ref{fig-position-times}, when the bubble is close to the solid seed, it will be easily and faster trapped by the solid dendrite. On the contrary, the bubble rising is mainly affected by the melt flow and is also pushed by the dendritic growth when it is located in the center space between two seeds. One can also observe that in the case of $x_0=Lx/8$, when there is a small channel formed by the symmetric two bubbles, the central dendrite is blocked by the bubble at first, and then grows quickly in the channel until recovers to a normal tip growth above the bubble. The influence of bubble on the dendritic growth can be displayed by the evolution of the tip velocity in Fig. \ref{fig-position-ComV}, where a little effect on tip velocity can be found in the cases when the bubble is away from the dendrite axis ($x_0=Lx/4$, $3Lx/8$, and $Lx/2$). However, the dynamics of bubble shows that there are some obvious differences due to the solid dendritic growth. As seen from Figs. \ref{fig-position-times} and \ref{fig-position-ComP}, the gas bubble rises and moves to the axis $x=0$ due to the symmetric vortexes caused by the gravity when it is initially placed at the asymmetric position ($x_0=Lx/8$, $Lx/4$, and $3Lx/8$). Furthermore, the bubble at $x_0=Lx/8$ is trapped quickly, while those at $x_0=Lx/4$ and $3Lx/8$ are pushed away from the central axis $x=0$ by the solid dendrite until they may be trapped by the side of dendrite. The effect of dendrite to bubble in the vertical direction is also represented, as shown in Fig. \ref{fig-position-ComP}(b).  


\subsection{The effect of seed number}
\begin{figure}
	\centering
	\subfigure[$n=2$]{
		\begin{minipage}{0.24\linewidth}
			\centering
			\includegraphics[width=1.6in]{fig-n2-4000.pdf}
	\end{minipage}}
	\subfigure[$n=3$]{
		\begin{minipage}{0.24\linewidth}
			\centering
			\includegraphics[width=1.6in]{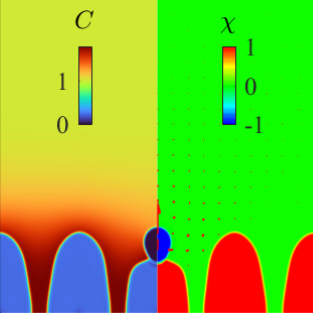}
	\end{minipage}}
	\subfigure[$n=4$]{
		\begin{minipage}{0.24\linewidth}
			\centering
			\includegraphics[width=1.6in]{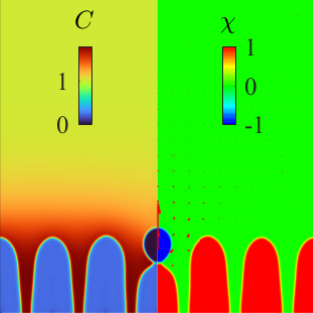}
	\end{minipage}}
	
	\subfigure[$n=2$]{
		\begin{minipage}{0.24\linewidth}
			\centering
			\includegraphics[width=1.6in]{fig-n2-10000.pdf}
	\end{minipage}}
	\subfigure[$n=3$]{
		\begin{minipage}{0.24\linewidth}
			\centering
			\includegraphics[width=1.6in]{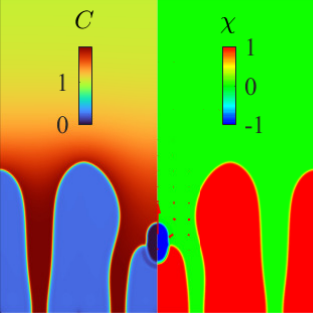}
	\end{minipage}}
	\subfigure[$n=4$]{
		\begin{minipage}{0.24\linewidth}
			\centering
			\includegraphics[width=1.6in]{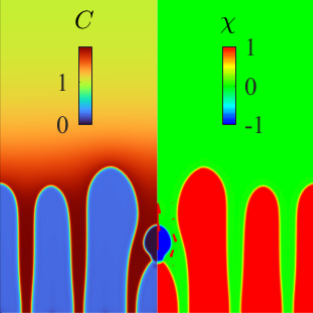}
	\end{minipage}}
	\caption{The snapshots of concentration and phase fields superimposed with velocity field with different numbers of solid seeds [$t/\tau_0=4000$ (top row) and $t/\tau_0=10000$ (bottom row)].}
	\label{fig-number}
\end{figure}

Finally, the effect of the number of initial solid seeds is discussed, and some snapshots of the concentration and phase fields with the solid seed number $n=2$, 3 and 4 are plotted in Fig. \ref{fig-number}. From this figure, one can find that with the increase of the seed number, the dendrites grow finer and the contact area with the bubble becomes smaller, such that the bubble cannot been totally trapped by the solid dendrite and has a significant deformation under the action of gravity and the pinning effect of solid dendrite.   

\section{Conclusions}\label{sec-conclusion}
In this paper, a phase-field model is proposed for the binary alloy solidification in presence of gas bubbles by introducing a general total free energy. In the present model, the general total free energy includes the energy for gas-liquid two-phase flows, the part for binary alloy solidification, and the interaction energy between solid dendrite and gas bubble. In the region away from the solid dendrite, the total energy reduces to the popular energy for gas-liquid flows, while it degenerates to the energy for thermosolutal dendritic growth in the solidification region. The phase-field equations can be obtained through minimizing the total free energy, and the conservative Allen-Cahn equation is used to replace the Cahn-Hilliard equation for gas-liquid interface capturing. To preserve the local solute vacuum in the gas bubbles, a flux term is added in the solute transport equation. In addition, some interpolation functions and parameters in the governing equations for the dendritic growth of binary alloy can be determined by the asymptotic analysis of thin-interface limit, and the present general form is identical to the previous model in a specific case. Similar to our previous work \cite{Zhan2023CiCP}, an extra force limited by the volume fraction of the solid phase is introduced to depict the fluid-solid interaction, which avoids the treatment of velocity boundary conditions on the complex fluid-solid interface. Then the present phase-field model is solved by the LBM, and its accuracy is tested by some benchmark problems. Finally, a complex problem of binary alloy solidification in presence of gas bubbles is considered, and the effects of some physical parameters on the interaction between solid dendrites and gas bubble are discussed. In the future, the present model will be extended to study the interactions of gas, liquid and solid in a welding molten pool.

\section*{Acknowledgments}
This research was supported by the National Natural Science Foundation of China (Grants No. 12072127 and No. 123B2018), the Interdisciplinary Research Program of HUST (2024JCYJ001 and 2023JCYJ002), and the Fundamental Research Funds for the Central Universities, HUST (No. YCJJ20241101 and No. 2023JY-CXJJ046). The computation was completed on the HPC Platform of Huazhong University of Science and Technology.

\appendix
\section{Asymptotic analysis of the thin-interface limit}\label{sec-analysis}
The goal of the matched asymptotic analysis is to relate the phase-field equations for alloy solidification [Eq. (\ref{eq-PFE}) with $\phi=\phi_l$] to the following free boundary problem,
\begin{subequations}
	\begin{equation}
		\partial_tU=D\nabla U\quad\text{(liquid)},
	\end{equation}
	\begin{equation}
		\partial_t\theta=\alpha\nabla^2\theta\quad\text{(liquid and solid)},
	\end{equation}
	\begin{equation}
		\left[1+\left(1-k\right)U_i\right]v_n=-D\partial_nU|_l\quad\text{(mass conservation)},
	\end{equation}
	\begin{equation}
		v_n=\alpha\left(\partial_n\theta|_s-\partial_n\theta|_l\right)\quad\text{(heat conservation)},
	\end{equation}
	\begin{equation}
		\theta_i+Mc_{\infty}U_i=-d_0\kappa_{\psi}-\beta v_n\quad\text{(Gibbs-Thomson)},
	\end{equation}
\end{subequations}
where $v_n$ is the local normal velocity of the interface, $d_0=\gamma T_Mc_p/L^2$ is the capillary length with $\gamma$ being the liquid-solid surface energy, $\kappa_{\psi}$ is the curvature, $\beta=c_p/\mu_kL$ is the kinetic coefficient with $\mu_k$ being the interface mobility.

We first rewrite the phase-field equation (\ref{eq-PFE}) with $\phi=\phi_l$ in a dimensionless form by introducing the length unit $l_c$, time unit $l_c^2/D$, and a small parameter $\epsilon=W_{\psi}/l_c$, 
\begin{subequations}
	\begin{equation}
		\bar{D}\epsilon^2\partial_t\psi=\epsilon^2\nabla^2\psi-f'_2\left(\psi,T_M\right)-\lambda p'\left(\psi\right)\left(\theta+Mc_{\infty}U\right),
	\end{equation}
	\begin{equation}
		\left[1-\left(1-k\right)h\left(\psi\right)\right]\partial_tU=\nabla\cdot\left\{q\left(\psi\right)\nabla U-\epsilon a\left(\psi\right)\left[1+\left(1-k\right)U\right]\partial_t\psi\mathbf{n}_{\psi}\right\}+\left[1+\left(1-k\right)U\right]\partial_th\left(\psi\right),
	\end{equation} 
	\begin{equation}
		\partial_t\theta=Le\nabla^2\theta+\partial_th\left(\psi\right),
	\end{equation}
\end{subequations}
where $\bar{D}=D\tau/W_{\psi}^2$, some material properties (the density $\rho$, specific heat per unit volume $c_p$) in liquid and solid phases are assumed to be uniform.
 
Then we expand the inner solutions in powers of $\epsilon$ as
\begin{equation}\label{eq-InnerEx}
	\psi=\psi_0+\epsilon\psi_1+\epsilon^2\psi_2+\cdots,\quad U=U_0+\epsilon U_1+\epsilon^2 U_2+\cdots,\quad \theta=\theta_0+\epsilon\theta_1+\epsilon^2\theta_2+\cdots,
\end{equation}
and similarly, in the outer region, $\tilde{\psi}=\tilde{\psi}_0+\epsilon\tilde{\psi}_1+\epsilon^2\tilde{\psi}_2+\cdots$, $\tilde{U}=\tilde{U}_0+\epsilon\tilde{U}_1+\epsilon^2\tilde{U}_2+\cdots$, $\tilde{\theta}=\tilde{\theta}_0+\epsilon\tilde{\theta}_1+\epsilon\tilde{\theta}_2+\cdots$. 
Since $\tilde{\psi}$ is constant in each phase, $\tilde{U}$ and $\tilde{\theta}$ simply obey the following diffusion equations,
\begin{equation}
	\partial_t\tilde{U}=\nabla\cdot q\left(\tilde{\psi}\right)\nabla\tilde{U},\quad \partial_t\tilde{\theta}=Le\nabla^2\tilde{\theta}.
\end{equation}
In the inner region, we introduce a local orthogonal set of curvilinear coordinates $\left(r,s\right)$ that moves with the local normal velocity of the liquid-solid interface, $r$ and $s$ measure the signed distance along the normal direction and the arclength along interface. Furthermore, the inner variable $\eta=r/\epsilon$, the dimensionless interface velocity $V_n=v_nl_c/D$ and the dimensionless curvature $\mathcal{K}=l_c\kappa_{\psi}$ are introduced such that the standard formulas of differential geometry yield \cite{Echebarria2004PRE,Folch1999PRE}
\begin{equation}
	\begin{aligned}
		\partial_t&=-\epsilon^{-1}V_n\partial_{\eta}+d_t-V_t\partial_s+O\left(\epsilon\right),\\
		\nabla^2&=\epsilon^{-2}\partial_{\eta\eta}+\epsilon^{-1}\mathcal{K}\partial_{\eta}-\mathcal{K}^2\eta\partial_{\eta}+\partial_{ss}+O\left(\epsilon\right),\\
		\nabla\cdot\left(q\nabla\right)&=\epsilon^{-2}\partial_{\eta}\left(q\partial_{\eta}\right)+\epsilon^{-1}\mathcal{K}q\partial_{\eta}-\mathcal{K}^2q\eta\partial_{\eta}+\partial_s\left(q\partial_s\right)+O\left(\epsilon\right),\\
		\mathbf{n}_{\psi}&=\mathbf{n}\left[1+O\left(\epsilon^2\right)\right]+\mathbf{s}O\left(\epsilon\right),\\
		\nabla\cdot\mathbf{a}&=\epsilon^{-1}\partial_{\eta}\left(\mathbf{n}\cdot\mathbf{a}\right)+\partial_s\left(\mathbf{s}\cdot\mathbf{a}\right)+\mathcal{K}\mathbf{n}\cdot\mathbf{a}+O\left(\epsilon\right),
	\end{aligned}
\end{equation}
where $V_t$ is the dimensionless tangential velocity of the interface, $q$ and $\mathbf{a}$ are the arbitrary functions, $\mathbf{n}$ is the unit normal vector and $\mathbf{s}$ is the tangential vector. 
Applying above formulas to the phase-field equations, we have
\begin{subequations}\label{eq-PFErs}
	\begin{equation}
		\epsilon\left(\bar{D}V_n+\mathcal{K}\right)\partial_{\eta}\psi+\partial_{\eta\eta}\psi+\epsilon^2\left(\partial_{ss}\psi-\mathcal{K}^2\eta\partial_{\eta}\psi\right)-f'_2\left(\psi,T_M\right)-\lambda p'\left(\psi\right)\left(\theta+Mc_{\infty}U\right)=0,
	\end{equation}
	\begin{equation}
		\begin{aligned}
			&\epsilon V_n\left[1-\left(1-k\right)h\left(\psi\right)\right]\partial_{\eta}U+\partial_{\eta}\left[q\left(\psi\right)\partial_{\eta}U\right]+\epsilon\mathcal{K}q\left(\psi\right)\partial_{\eta}U-\epsilon^2\mathcal{K}^2q\left(\psi\right)\eta\partial_{\eta}U+\epsilon^2\partial_s\left[q\left(\psi\right)\partial_sU\right]\\
			&+\epsilon V_n\partial_{\eta}\left(a\left(\psi\right)\left[1+\left(1-k\right)U\right]\partial_{\eta}\psi\right)+\epsilon^2a\left(\psi\right)V_n\mathcal{K}\left[1+\left(1-k\right)U\right]\partial_{\eta}\psi-\epsilon V_n\left[1+\left(1-k\right)U\right]\partial_{\eta}h\left(\psi\right)=0,
		\end{aligned}
	\end{equation}
	\begin{equation}
		\epsilon\left(V_n+Le\mathcal{K}\right)\partial_{\eta}\theta+Le\partial_{\eta\eta}\theta+\epsilon^2Le\left(\partial_{ss}\theta-\mathcal{K}^2\eta\partial_{\eta}\theta\right)-\epsilon V_n\partial_{\eta}h\left(\psi\right)=0,
	\end{equation}
\end{subequations}
where the terms in $V_t$ and $d_t$ and higher-order terms in $\epsilon^3$ are neglected.

Substituting the inner expansion (\ref{eq-InnerEx}) into Eq. (\ref{eq-PFErs}), one can obtain equations at different orders of $\epsilon$. At the leading order of $\epsilon$, we have 
\begin{equation}
	\partial_{\eta\eta}\psi_0-f'_2\left(\psi_0,T_M\right)-\lambda p'\left(\psi_0\right)\left(\theta_0+Mc_{\infty}U_0\right)=0,\quad \partial_{\eta}\left[q\left(\psi_0\right)\partial_{\eta}U_0\right]=0,\quad \partial_{\eta\eta}\theta_0=0,
\end{equation}
from which one can derive the trivial solutions, $\theta_0+Mc_{\infty}U_0=0$ with $\theta_0$ and $U_0$ being constants, and
\begin{equation}
	\psi_0=\frac{\psi_s+\psi_l}{2}-\frac{\psi_s-\psi_l}{2}\tanh\left(\frac{\psi_s-\psi_l}{2}\sqrt{2\beta_{\psi}}\eta\right),
\end{equation}
where $\eta\rightarrow+\infty$, $\psi_0\rightarrow\psi_l$ in liquid phase, while $\eta\rightarrow-\infty$, $\psi_0\rightarrow\psi_s$ in solid phase.

At the first order of $\epsilon$, we can obtain
\begin{subequations}
	\begin{equation}
		\mathcal{L}\psi_1=\lambda p'\left(\psi_0\right)\left(\theta_1+Mc_{\infty}U_1\right)-\left(\bar{D}V_n+\mathcal{K}\right)\partial_{\eta}\psi_0=0,
	\end{equation}
	\begin{equation}\label{eq-CO1}
		\partial_{\eta}\left[q\left(\psi_0\right)\partial_{\eta}U_1\right]-V_n\left[1+\left(1-k\right)U_0\right]\partial_{\eta}\left[h\left(\psi_0\right)-a\left(\psi_0\right)\partial_{\eta}\psi_0\right]=0,
	\end{equation}
	\begin{equation}\label{eq-TO1}
		\partial_{\eta}\left[Le\partial_{\eta}\theta_1-V_nh\left(\psi_0\right)\right]=0,
	\end{equation}
\end{subequations}
where $\mathcal{L}=\partial_{\eta\eta}-f''_2\left(\psi_0,T_M\right)$ is a linear operator. Integrating Eqs (\ref{eq-CO1}) and (\ref{eq-TO1}) directly yields
\begin{equation}
	q\left(\psi_0\right)\partial_{\eta}U_1=A_1+V_n\left[1+\left(1-k\right)U_0\right]\left[h\left(\psi_0\right)-a\left(\psi_0\right)\partial_{\eta}\psi_0\right],\quad \partial_{\eta}\theta_1=B_1+\frac{V_n}{Le}h\left(\psi_0\right).
\end{equation}
When $\eta\rightarrow-\infty$, we have $\psi_0\rightarrow\psi_s$, $q\left(\psi_0\right)\rightarrow0$, $h\left(\psi_0\right)\rightarrow1$ and $\partial_{\eta}\psi_0\rightarrow0$, which lead to $A_1=-V_n\left[1+\left(1-k\right)U_0\right]$. Then the following equations can be derived,
\begin{equation}\label{eq-CTint1}
	\partial_{\eta}U_1=V_n\left[1+\left(1-k\right)U_0\right]g\left(\psi_0\right),\quad \partial_{\eta}\theta_1=B_1+\frac{V_n}{Le}h\left(\psi_0\right),\quad g\left(\psi_0\right)=\frac{h\left(\psi_0\right)-a\left(\psi_0\right)\partial_{\eta}\psi_0-1}{q\left(\psi_0\right)}.
\end{equation}
Integrating the above equation once more yields
\begin{equation}
	U_1=\bar{U}_1+V\left[1+\left(1-k\right)U_0\right]\int_{0}^{\eta}g\left(\psi_0\right)\mathrm{d}\zeta,\quad \theta_1=\bar{\theta}_1+B_1\eta+\frac{V}{Le}\int_{0}^{\eta}h\left(\psi_0\right)\mathrm{d}\zeta,
\end{equation}
where $\bar{U}_1$ and $\bar{\theta}_1$ are the values of $U_1$ and $\theta_1$ at the interface. The profiles of $U_1$ and $\theta_1$ depend on the choice of the function $a\left(\psi\right)$. It should be noted that when $\eta\rightarrow-\infty$, both the denominator and the numerator would approach to zero. Here it is important to remark that we need to require $g\left(\psi\right)\rightarrow0$ in this limit, i.e., the numerator must vanish more rapidly than the denominator, otherwise $U_1$ would diverge. With the help of the definitions of $h\left(\psi\right)$ and $q\left(\psi\right)$, and $\partial_{\eta}\psi_0=-\sqrt{2\beta_{\psi}}\left(\psi_s-\psi\right)\left(\psi-\psi_l\right)$, one can obtain
\begin{equation}
	g\left(\psi_0\right)=-1+a\left(\psi_0\right)\sqrt{2\beta_{\psi}}\left(\psi-\psi_l\right)\left(\psi_s-\psi_l\right).
\end{equation}
If $a\left(\psi_0\right)=1/\left[\sqrt{2\beta_{\psi}}\left(\psi_s-\psi_l\right)^2\right]$, we have $g\left(\psi_0\right)=-1+h\left(\psi_0\right)$, and then
\begin{equation}\label{eq-CTint2}
	U_1=\bar{U}_1+V_n\left[1+\left(1-k\right)U_0\right]\left[-\eta+\int_{0}^{\eta}h\left(\psi_0\right)\mathrm{d}\zeta\right],\quad \theta_1=\bar{\theta}_1+B_1\eta+\frac{V}{Le}\int_{0}^{\eta}h\left(\psi_0\right)\mathrm{d}\zeta.
\end{equation}

Now we focus on the matching conditions between two expansions at the interface region. The condition that slopes of the two solutions match on the solid and liquid sides of the interface in the region defined by $1\ll|\eta|\ll\epsilon^{-1}$ implies that
\begin{equation}\label{eq-match}
	\lim_{\eta\rightarrow\pm\infty}\partial_{\eta}U_1=\lim_{r\rightarrow0^\pm}\partial_{r}\tilde{U}_0=\partial_{r}\tilde{U}_0|^\pm,\, \partial_{r}\tilde{U}_1|^\pm=\lim_{\eta\rightarrow\pm\infty}\left[\partial_{\eta}U_2-\eta\partial_{rr}\tilde{U}_0|^\pm\right],\, \lim_{\eta\rightarrow\pm\infty}\partial_{\eta}\theta_1=\lim_{r\rightarrow0^\pm}\partial_{r}\tilde{\theta}_0=\partial_{r}\tilde{\theta_0}|^\pm.
\end{equation}
Applying the above matching conditions to Eq. (\ref{eq-CTint1}), we have
\begin{subequations}
	\begin{equation}
		\partial_{r}\tilde{U}_0|^-=0,\quad \partial_{r}\tilde{U}_0|^+=-V_n\left[1+\left(1-k\right)U_0\right],
	\end{equation} 
	\begin{equation}
		\partial_{r}\tilde{\theta}_0|^-=B_1+\frac{V_n}{Le},\quad \partial_{r}\tilde{\theta}_0|^+=B_1.
	\end{equation}
\end{subequations}
Thus the following conditions can be derived,
\begin{equation}\label{eq-Stefan}
	V_n\left[1+\left(1-k\right)U_0\right]=-\partial_{r}\tilde{U}_0|^+,\quad V_n=Le\left(\partial_{r}\tilde{\theta}_0|^--\partial_{r}\tilde{\theta}_0|^+\right),
\end{equation}
which is just the Stefan condition at the lowest order. To evaluate eventual corrections, the terms at order $\epsilon$ should be considered, and the process is similar to that in Ref. \cite{Echebarria2004PRE}. The outer problem satisfies the diffusion equations in a moving curvilinear coordinate system, therefore one can obtain $\left[\partial_{rr}+\left(V_n+\mathcal{K}\right)\partial_{r}+\partial_{ss}\right]\tilde{U}_0=0$ and $\left[Le\partial_{rr}+\left(V_n+Le\mathcal{K}\right)\partial_{r}+\partial_{ss}\right]\tilde{\theta}_0=0$, such that $\partial_{rr}\tilde{U}_0|^\pm=-\left[\left(V_n+\mathcal{K}\right)\partial_{r}+\partial_{ss}\right]\tilde{U}_0|^\pm$ and $\partial_{rr}\tilde{\theta}_0|^\pm=-\left[\left(V_n/Le+\mathcal{K}\right)\partial_{r}+\partial_{ss}\right]\tilde{\theta}_0|^\pm$.

At the second order of $\epsilon$, we can obtain the following formulas from Eq. (\ref{eq-PFErs}),
\begin{equation}
	\begin{aligned}
		&V_n\left[1-\left(1-k\right)h\left(\psi_0\right)\right]\partial_{\eta}U_1+\mathcal{K}q\left(\psi_0\right)\partial_{\eta}U_1-V_n\left(1-k\right)U_1h'\left(\psi_0\right)\partial_{\eta}\psi_0-V_n\left[1+\left(1-k\right)U_0\right]\left[h'\left(\psi_0\right)\partial_{\eta}\psi_1+\psi_1\partial_{\eta}h'\left(\psi_0\right)\right]\\
		&+V_n\partial_{\eta}\left\{a'\left(\psi_0\right)\psi_1\left[1+\left(1-k\right)U_0\right]\partial_{\eta}\psi_0+a\left(\psi_0\right)\left(1-k\right)U_1\partial_{\eta}\psi_0+a\left(\psi_0\right)\left[1+\left(1-k\right)U_0\right]\partial_{\eta}\psi_1\right\}\\
		&+\partial_{\eta}\left[q\left(\psi_0\right)\partial_{\eta}U_2+q'\left(\phi_l,\psi_0\right)\psi_1\partial_{\eta}U_1\right]+a\left(\psi_0\right)V\mathcal{K}\left[1+\left(1-k\right)U_0\right]\partial_{\eta}\psi_0+q\left(\psi_0\right)\partial_{ss}U_0=0,
	\end{aligned}
\end{equation}
and
\begin{equation}
	\left(V_n+Le\mathcal{K}\right)\partial_{\eta}\theta_1+Le\partial_{\eta\eta}\theta_2+Le\partial_{ss}\theta_0-V_n\partial_{\eta}\left(h'\left(\psi_0\right)\psi_1\right)=0,
\end{equation}
where $\partial_{\eta}U_0=\partial_{\eta}\theta_0=\partial_s\psi_0=0$ has been used. Integrating above equations from 0 to $\eta$ yields
\begin{equation}\label{eq-2rdCint}
	\begin{aligned}
		&V_n\int_{0}^{\eta}\left[1-\left(1-k\right)h\left(\psi_0\right)\right]\partial_{\zeta}U_1\mathrm{d}\zeta+\mathcal{K}\int_{0}^{\eta}q\left(\psi_0\right)\partial_{\zeta}U_1\mathrm{d}\zeta-V_n\left(1-k\right)\int_{0}^{\eta}U_1\partial_{\zeta}h\left(\psi_0\right)\mathrm{d}\zeta-V_n\left[1+\left(1-k\right)U_0\right]h'\left(\psi_0\right)\psi_1\\
		&+V_n\left\{a'\left(\psi_0\right)\psi_1\left[1+\left(1-k\right)U_0\right]\partial_{\eta}\psi_0+a\left(\psi_0\right)\left(1-k\right)U_1\partial_{\eta}\psi_0+a\left(\psi_0\right)\left[1+\left(1-k\right)U_0\right]\partial_{\eta}\psi_1\right\}\\
		&+q\left(\psi_0\right)\partial_{\eta}U_2+q'\left(\phi_l,\psi_0\right)\psi_1\partial_{\eta}U_1+V\mathcal{K}\left[1+\left(1-k\right)U_0\right]\int_{0}^{\eta}a\left(\psi_0\right)\partial_{\eta}\psi_0\mathrm{d}\zeta+\partial_{ss}U_0\int_{0}^{\eta}q\left(\psi_0\right)\mathrm{d}\zeta=A_2,
	\end{aligned}
\end{equation}
and
\begin{equation}
	\left(V_n+Le\mathcal{K}\right)\theta_1+Le\partial_{\eta}\theta_2+Le\eta\partial_{ss}\theta_0-Vh'\left(\psi_0\right)\psi_1=B_2.
\end{equation}
Actually, we are only interested in the results at the limit $\eta\rightarrow\pm\infty$. In this limit, $\phi_1$ and $\partial_{\eta}\psi_0$ are exponentially small, thus the terms containing them can be dropped, except when they appear in an integral. Some remaining piece can be rewritten as
\begin{equation}
	V_n\int_{0}^{\eta}\left[1-\left(1-k\right)h\left(\psi_0\right)\right]\partial_{\zeta}U_1\mathrm{d}\zeta-V_n\left(1-k\right)\int_{0}^{\eta}U_1\partial_{\zeta}h\left(\psi_0\right)\mathrm{d}\zeta=V_n\left[1-\left(1-k\right)h\left(\psi_0\right)\right]U_1.
\end{equation}
In addition, according to Eqs. (\ref{eq-CTint1}) and (\ref{eq-Stefan}), we have 
\begin{equation}
	\begin{aligned}
		&\mathcal{K}\int_{0}^{\eta}q\left(\psi_0\right)\partial_{\zeta}U_1\mathrm{d}\zeta+V_n\mathcal{K}\left[1+\left(1-k\right)U_0\right]\int_{0}^{\eta}a\left(\psi_0\right)\partial_{\eta}\psi_0\mathrm{d}\zeta=\mathcal{K}V_n\left[1+\left(1-k\right)U_0\right]\int_{0}^{\eta}\left[h\left(\psi_0\right)-1\right]\mathrm{d}\zeta\\
		&=\mathcal{K}\eta\partial_{r}\tilde{U}_0|^\pm+\mathcal{K}V\left[1+\left(1-k\right)U_0\right]\int_{0}^{\eta}\left[h\left(\psi_0\right)-h\left(\psi^\pm\right)\right]\mathrm{d}\zeta.
	\end{aligned}
\end{equation}
Using the matching condition for $U_1$, $\lim_{\eta\rightarrow\pm\infty}U_1=\tilde{U}_1|^\pm+\eta\partial_{r}\tilde{U}_0|^\pm$, and the fact $\lim_{\eta\rightarrow-\infty}q\left(\psi_0\right)\partial_{\eta}U_2=0$, we can obtain the constant $A_2$ from the limit $\eta\rightarrow-\infty$ of Eq. (\ref{eq-2rdCint}),
\begin{equation}
	\begin{aligned}
		A_2&=V_n\left[1-\left(1-k\right)h\left(\psi_0\right)\right]U_1+\mathcal{K}\eta\partial_{r}\tilde{U}_0|^-+\mathcal{K}V_n\left[1+\left(1-k\right)U_0\right]\int_{0}^{\eta}\left[h\left(\psi_0\right)-1\right]\mathrm{d}\zeta\\
		&=V_nk\left(\tilde{U}_1|^-+\eta\partial_{r}\tilde{U}_0|^-\right)+\mathcal{K}\eta\partial_{r}\tilde{U}_0|^-+\mathcal{K}V_n\left[1+\left(1-k\right)U_0\right]\int_{0}^{-\infty}\left[h\left(\psi_0\right)-1\right]\mathrm{d}\eta+\partial_{ss}U_0\int_{0}^{-\infty}q\left(\psi_0\right)\mathrm{d}\zeta.
	\end{aligned}
\end{equation}
Next, the term $\lim_{\eta\rightarrow+\infty}q\left(\psi_0\right)\partial_{\eta}U_2$ can be evaluated by
\begin{equation}
	\begin{aligned}
		\lim_{\eta\rightarrow+\infty}q\left(\psi_0\right)\partial_{\eta}U_2=&A_2-V_n\left[1-\left(1-k\right)h\left(\psi_0\right)\right]U_1-\mathcal{K}\eta\partial_{r}\tilde{U}_0|^+-\mathcal{K}V_n\left[1+\left(1-k\right)U_0\right]\int_{0}^{\infty}h\left(\psi_0\right)\mathrm{d}\eta-\partial_{ss}U_0\int_{0}^{\infty}q\left(\psi_0\right)\mathrm{d}\zeta\\
		=&A_2-V_n\left(\tilde{U}_1|^++\eta\partial_{r}\tilde{U}_0|^+\right)-\mathcal{K}\eta\partial_{r}\tilde{U}_0|^+-\mathcal{K}V_n\left[1+\left(1-k\right)U_0\right]\int_{0}^{\infty}h\left(\psi_0\right)\mathrm{d}\eta-\partial_{ss}\tilde{U}_0\int_{0}^{\infty}q\left(\psi_0\right)\mathrm{d}\zeta\\
		=&-V_nk\left(\tilde{U}_1|^+-\tilde{U}_1|^-\right)-\left(1-k\right)V_n\tilde{U}_1|^+-\eta V_n\partial_{r}\tilde{U}_0|^+-\mathcal{K}\eta\partial_{r}\tilde{U}_0|^+\\
		&-\mathcal{K}V_n\left[1+\left(1-k\right)U_0\right]\left(H^+-H^-\right)-\partial_{ss}\tilde{U}_0\int_{0}^{\infty}q\left(\psi_0\right)\mathrm{d}\eta+\partial_{ss}\tilde{U}_0\int_{0}^{-\infty}q\left(\psi_0\right)\mathrm{d}\eta.
	\end{aligned}
\end{equation}
Finally, we have
\begin{equation}
	\begin{aligned}
		\partial_{r}\tilde{U}_1|^+&=\lim_{\eta\rightarrow+\infty}\left[\partial_{\eta}U_2-\eta\partial_{rr}\tilde{U}_0|^+\right]=\lim_{\eta\rightarrow+\infty}q\left(\psi_0\right)\partial_{\eta}U_2+\lim_{\eta\rightarrow+\infty}\eta\left[\left(V_n+\mathcal{K}\right)\partial_{r}+\partial_{ss}\right]\tilde{U}_0|^+\\
		&=\lim_{\eta\rightarrow+\infty}q\left(\psi_0\right)\partial_{\eta}U_2+V_n\lim_{\eta\rightarrow+\infty}\eta\partial_{r}\tilde{U}_0|^++\lim_{\eta\rightarrow+\infty}\mathcal{K}\eta\partial_{r}\tilde{U}_0|^++\lim_{\eta\rightarrow+\infty}\eta\partial_{ss}\tilde{U}_0\\
		&=-V_nk\left(\tilde{U}_1|^+-\tilde{U}_1|^-\right)-\left(1-k\right)V_n\tilde{U}_1|^+-\mathcal{K}V_n\left[1+\left(1-k\right)U_0\right]\left(H^+-H^-\right)-\partial_{ss}\tilde{U}_0\left(Q^+-Q^-\right),
	\end{aligned}
\end{equation}
and 
\begin{equation}
	\begin{aligned}
		\partial_{r}\tilde{\theta}_1|^\pm=&\lim_{\eta\rightarrow\pm\infty}\left[\partial_{\eta}\theta_2-\eta\partial_{rr}\tilde{\theta}_0|^\pm\right]=\frac{B_2}{Le}-\left(\frac{V_n}{Le}+\mathcal{K}\right)\lim_{\eta\rightarrow\pm\infty}\theta_1-\lim_{\eta\rightarrow\pm\infty}\eta\partial_{ss}\theta_0-\lim_{\eta\rightarrow\pm\infty}\eta\partial_{rr}\tilde{\theta}_0\\
		=&\frac{B_2}{Le}-\left(\frac{V_n}{Le}+\mathcal{K}\right)\lim_{\eta\rightarrow\pm\infty}\left(\bar{\theta}_1+B_1\eta+\frac{V_n}{Le}\int_{0}^{\eta}h\left(\psi_0\right)\mathrm{d}\zeta\right)+\lim_{\eta\rightarrow\pm\infty}\left(\frac{V_n}{Le}+\mathcal{K}\right)\eta\partial_{r}\tilde{\theta}_0|^\pm\\
		=&\frac{B_2}{Le}-\left(\frac{V_n}{Le}+\mathcal{K}\right)\bar{\theta}_1-\left(\frac{V_n}{Le}+\mathcal{K}\right)\left(\lim_{\eta\rightarrow\pm\infty}\eta\partial_{r}\tilde{\theta}_0|^\pm+\frac{V_n}{Le}H^\pm\right)+\lim_{\eta\rightarrow\pm\infty}\left(\frac{V_n}{Le}+\mathcal{K}\right)\eta\partial_{r}\tilde{\theta}_0|^\pm\\
		=&\frac{B_2}{Le}-\left(\frac{V_n}{Le}+\mathcal{K}\right)\left(\bar{\theta}_1+\frac{V_n}{Le}H^\pm\right),
	\end{aligned}
\end{equation}
where
\begin{equation}
	H^\pm=\int_{0}^{\pm\infty}\left[h\left(\psi_0\right)-h\left(\psi^\mp\right)\right]\mathrm{d}\eta,\quad Q^\pm=\int_{0}^{\pm\infty}\left[q\left(\psi_0\right)-q\left(\psi^\mp\right)\right]\mathrm{d}\eta.
\end{equation}
In addition, one can also obtain the following relation,
\begin{equation}
	\tilde{U}_1|^\pm=\lim_{\eta\rightarrow\pm\infty}U_1-\eta\partial_{r}\tilde{U}_0|^\pm=\bar{U}_0+V_n\left[1+\left(1-k\right)U_0\right]\lim_{\eta\rightarrow\pm\infty}\int_{0}^{\eta}g\left(\psi_0\right)\mathrm{d}\zeta-\eta\partial_{r}\tilde{U}_0|^\pm=\bar{U}_0+V_n\left[1+\left(1-k\right)U_0\right]H^\pm.
\end{equation}
Based on the definitions of $h\left(\psi_0\right)$ and $q\left(\psi_0\right)$, one can derive $H^+=H^-=H$ and $Q^+=Q^-=Q$.

In a word, we obtain the mass conservation condition and the heat conservation condition,
\begin{equation}
	\begin{aligned}
		&V_n\left[1+\left(1-k\right)U|^+\right]=V_n\left[1+\left(1-k\right)\left(\tilde{U}_0+\epsilon\tilde{U}_1|^+\right)\right]=-\partial_{r}\tilde{U}_0|^++\left(1-k\right)V_n\epsilon\tilde{U}_1|^+\\
		&=-\partial_{r}\left(\tilde{U}_0+\epsilon\tilde{U}_1|^+\right)-\epsilon \left[V_nk\left(\tilde{U}_1|^+-\tilde{U}_1|^-\right)+\mathcal{K}V_n\left[1+\left(1-k\right)U_0\right]\left(H^+-H^-\right)-\partial_{ss}\tilde{U}_0\left(Q^+-Q^-\right)\right]\\
		&=-\partial_{r}\left(\tilde{U}_0+\epsilon\tilde{U}_1|^+\right)-\epsilon \left[\left(V_n^2k+\mathcal{K}V_n\right)\left[1+\left(1-k\right)U_0\right]\left(H^+-H^-\right)-\partial_{ss}\tilde{U}_0\left(Q^+-Q^-\right)\right]\\
		&=-\partial_{r}U|^+,
	\end{aligned}
\end{equation}
and
\begin{equation}
	Le\left(\partial_{r}\tilde{\theta}|^--\partial_{r}\tilde{\theta}|^+\right)=Le\left(\partial_{r}\tilde{\theta}_0|^--\partial_{r}\tilde{\theta}_0|^+\right)+Le\epsilon\left(\partial_{r}\tilde{\theta}_1|^--\partial_{r}\tilde{\theta}_1|^+\right)=V_n.
\end{equation}

Now we differentiate the leading order phase-field equation with respect to $\eta$,
\begin{equation}
	\partial_{\eta\eta\eta}\psi_0-f''_2\left(\psi_0,T_M\right)\partial_{\eta}\psi_0=\mathcal{L}\partial_{\eta}\psi_0=0,
\end{equation}  
which implies that $\partial_{\eta}\psi_0$ is a homogeneous solution of $\mathcal{L}\partial_{\eta}\psi_0=0$. Then we have the following solvability condition,
\begin{equation}
	\begin{aligned}
		0=&\int_{-\infty}^{\infty}\partial_{\eta}\psi_0\left[\lambda p'\left(\psi_0\right)\left(\theta_1+Mc_{\infty}U_1\right)-\left(\bar{D}V_n+\mathcal{K}\right)\partial_{\eta}\psi_0\right]\mathrm{d}\eta\\
		=&\int_{-\infty}^{\infty}\lambda p'\left(\psi_0\right)\partial_{\eta}\psi_0\left\{\bar{\theta}_1+B_1\eta+\frac{V_n}{Le}\int_{0}^{\eta}h\left(\psi_0\right)\mathrm{d}\zeta+Mc_{\infty}\bar{U}_1+Mc_{\infty}V_n\left[1+\left(1-k\right)U_0\right]\int_{0}^{\eta}g\left(\psi_0\right)\mathrm{d}\zeta\right\}\mathrm{d}\eta\\
		&-\int_{-\infty}^{\infty}\left(\bar{D}V_n+\mathcal{K}\right)\left(\partial_{\eta}\psi_0\right)^2\mathrm{d}\eta,
	\end{aligned}
\end{equation}
from which we get
\begin{equation}\label{eq-BarCT}
	\bar{\theta}_1+Mc_{\infty}\bar{U}_1=-\frac{I}{\lambda J}\left(\bar{D}V_n+\mathcal{K}\right)+\frac{BF}{J}+\frac{V_nK}{J}\left\{\frac{1}{Le}+\left[1+\left(1-k\right)U_0\right]Mc_{\infty}\right\},
\end{equation}
where $B=B_1+\left[1+\left(1-k\right)U_0\right]Mc_{\infty}$, 
\begin{equation}
	\begin{aligned}
		I&=\int_{-\infty}^{+\infty}\left(\partial_{\eta}\psi_0\right)^2\mathrm{d}\eta,\quad J=-\int_{-\infty}^{+\infty}p'\left(\psi_0\right)\partial_{\eta}\psi_0\mathrm{d}\eta,\\
		F&=\int_{-\infty}^{+\infty}\eta p'\left(\psi_0\right)\partial_{\eta}\psi_0\mathrm{d}\eta,\quad K=\int_{-\infty}^{+\infty}p'\left(\psi_0\right)\partial_{\eta}\psi_0\left(\int_{0}^{\eta}h\left(\psi_0\right)\mathrm{d}\zeta\right)\mathrm{d}\eta.
	\end{aligned}
\end{equation}

To determine the conditions of the outer solution on the two sides of the interface, we expand $\tilde{\theta}$ and $\tilde{U}$ in the matching regions in terms of $r$,
\begin{equation}\label{eq-OuterI}
	\tilde{\theta}+Mc_{\infty}\tilde{U}=\theta_i^\pm+Mc_{\infty}U_i^\pm+\left(\partial_{r}\tilde{\theta}_0|^\pm+Mc_{\infty}\partial_{r}\tilde{U}_0|^\pm\right)r.
\end{equation}
In the matching regions on both sides of the interface, the inner solution takes the form
\begin{equation}\label{eq-InnerI}
	\theta+Mc_{\infty}U=\epsilon\left[\bar{\theta}_1+Mc_{\infty}\bar{U}_1+\frac{V_n}{Le}H+Mc_{\infty}V_n\left[1+\left(1-k\right)U_0\right]H\right]+r\left(\partial_{r}\tilde{\theta}_0|^\pm+Mc_{\infty}\partial_{r}\tilde{U}_0|^\pm\right).
\end{equation}
Equating the terms on the right-hand side of Eqs. (\ref{eq-OuterI}) and (\ref{eq-InnerI}), we obtain the desired relation
\begin{equation}\label{eq-InterCT}
	\theta_i^\pm+Mc_{\infty}U_i^\pm=\epsilon\left[\bar{\theta}_1+Mc_{\infty}\bar{U}_1+\frac{V_n}{Le}H+Mc_{\infty}V_n\left[1+\left(1-k\right)U_0\right]H\right].
\end{equation} 
Then combing Eqs. (\ref{eq-BarCT}) and (\ref{eq-InterCT}) yields
\begin{equation}
	\begin{aligned}
		\theta_i+Mc_{\infty}U_i=&\epsilon\left\{-\frac{I}{\lambda J}\left(\bar{D}V_n+\mathcal{K}\right)+\frac{BF}{J}+\frac{V_nK}{LeJ}+\frac{V_nK}{J}\left[1+\left(1-k\right)U_0\right]Mc_{\infty}\right\}\\
		&+\epsilon\left\{\frac{V_n}{Le}H+Mc_{\infty}V_n\left[1+\left(1-k\right)U_0\right]H\right\}\\
		=&-a_1\frac{W}{\lambda}\kappa_{\psi}-\nu_na_1\left[\frac{\tau}{\lambda W}-\frac{a_2W}{D}\left(\frac{1}{Le}+Mc_{\infty}\left[1+\left(1-k\right)U_0\right]\right)\right],
	\end{aligned}
\end{equation}
where $a_1=I/J$ and $a_2=\left(K+JH\right)/I$ with
\begin{equation}
	I=\frac{\sqrt{2\beta_{\psi}}}{6}\left(\psi_s-\psi_l\right)^3,\quad J=1,\quad F=0,\quad H=\frac{\ln2}{\sqrt{2\beta_{\psi}}\left(\psi_s-\psi_l\right)},\quad K=\frac{47/60-\ln2}{\sqrt{2\beta_{\psi}}\left(\psi_s-\psi_l\right)}.
\end{equation}
Thus the Gibbs-Thomson condition can be recovered,
\begin{equation}
	\theta_i+Mc_{\infty}U_i=-d_0\kappa_{\psi}-\beta v_n,
\end{equation}
where
\begin{equation}
	d_0=a_1\frac{W}{\lambda},\quad\beta=a_1\left[\frac{\tau}{\lambda W}-\frac{a_2W}{D}\left(\frac{1}{Le}+Mc_{\infty}\left[1+\left(1-k\right)U_0\right]\right)\right].
\end{equation}
Additionally, for the low-speed solidification where the kinetic coefficient $\beta$ vanishes, one can obtain
\begin{equation}
	\tau=\tau_0\left(\frac{1}{Le}+Mc_{\infty}\left[1+\left(1-k\right)U_0\right]\right),\quad \tau_0=a_2\frac{\lambda W^2}{D}.
\end{equation}

Finally, it is worth noting that if we introduce $\lambda'=15\lambda/16$, under the condition of $d_0/W=a_1/\lambda=a_1'/\lambda'$ and $\tau_0 D/W^2=a_2\lambda=a_2'\lambda'$, one can obtain $a_1'=a_1\lambda'/\lambda=5\sqrt{2}/8$ and $a_2'=a_2\lambda/\lambda'=47/75$ for the case of $\psi_s=1$ and $\psi_l=-1$, which is identical to those in Ref. \cite{Karma1998PRE}.

\section{Lattice Boltzmann models for different physical fields}\label{sec-LBE}
In this appendix, we will present some specific lattice Boltzmann models for the physical fields considered in the work. Since the lattice Boltzmann model for the conservative Allen-Cahn equation is the same as those in some available works \cite{Zhan2022PRE,Wang2019Capillarity}, the model for the concentration and Navier-Stokes equations are consistent with those in our previous works \cite{Zhan2023CiCP,Zhan2022PRE,Zhan2024PD}, the details on these lattice Boltzmann models are not shown here. However, the anisotropic Allen-Cahn equation and the temperature equation for dendritic growth are more general than those in Ref. \cite{Zhan2023CiCP}, and the lattice Boltzmann models for these equations will be presented below.

When $Le=O\left(1\right)$, we can directly use the previous scheme in Ref. \cite{Zhan2023CiCP} and set $a_n=a_s^2\left(\mathbf{n}_{\psi}\right)F\left(U\right)$ in the lattice Boltzmann model for the anisotropic Allen-Cahn equation (\ref{eq-AnACE}). However, to avoid the numerical instability caused by a small or large $Le$ in $F\left(U\right)$, we rewrite Eq. (\ref{eq-AnACE}) as
\begin{equation}
	a_s^2\left(\mathbf{n}_{\psi}\right)\frac{\partial\psi}{\partial t}=\nabla\cdot\frac{W_0^2a_s^2\left(\mathbf{n}_{\psi}\right)}{\tau_0F\left(U\right)}\left[\nabla\psi+\frac{\mathbf{N}}{a_s^2\left(\mathbf{n}_{\psi}\right)}\right]+\frac{W_0^2}{\tau_0F^2\left(U\right)}\nabla F\left(U\right)\cdot\left[a_s^2\left(\mathbf{n}_{\psi}\right)\nabla\psi+\mathbf{N}\right]+\frac{Q_{\psi}}{\tau_0F\left(U\right)},
\end{equation}   
where $Q_{\psi}=-\left[f'_2\left(\psi,T_M\right)+\partial_{\psi}f_3\left(\phi,\psi\right)+\lambda p'\left(\psi\right)\left(\theta+Mc_{\infty}U\right)\right]$.
Similar to that in Ref. \cite{Zhan2023CiCP}, the lattice Boltzmann model for the above equation can be given by
\begin{equation}
	f_i\left(\mathbf{x}+\mathbf{c}_i\Delta t,t+a_s^2\left(\mathbf{n}_{\psi}\right)\Delta t\right)=f_i\left(\mathbf{x},t\right)-\Lambda_{ij}^{\psi}\left(f_j-f_j^{eq}\right)\left(\mathbf{x},t\right)+\Delta tF_i^{\psi}\left(\mathbf{x},t\right)+\frac{\Delta t^2}{2}\partial_tF_i^{\psi}\left(\mathbf{x},t\right)+\Delta t\left(\delta_{ij}-\frac{\Lambda_{ij}^{\psi}}{2}\right)G_j^{\psi}\left(\mathbf{x},t\right),
\end{equation}
where the distribution functions are designed as
\begin{equation}
	f_i^{eq}=\omega_i\psi,\quad F_i^{\psi}=\frac{\omega_iW_0^2}{\tau_0F^2\left(U\right)}\nabla F\left(U\right)\cdot\left[a_s^2\left(\mathbf{n}_{\psi}\right)\nabla\psi+\mathbf{N}\right]+\frac{\omega_iQ_{\psi}}{\tau_0F\left(U\right)},\quad G_i^{\psi}=-\frac{\omega_i\mathbf{c}_i\cdot\mathbf{N}}{a_s^2\left(\mathbf{n}_{\psi}\right)}.
\end{equation}
Additionally, the macroscopic order parameter $\psi$ can be calculated by
\begin{equation}
	\psi=\sum_if_i, 
\end{equation}
and the corresponding relaxation parameter $s_1^{\psi}$ is determined by the relation  $W_0^2a_s^2\left(\mathbf{n}_{\psi}\right)/\tau_0F\left(U\right)=\left(1/s_1^{\psi}-1/2\right)c_s^2\Delta t$.

For the temperature field, through including a variable $H=c_p\theta-c_p^sh\left(\psi\right)$, the temperature equation can be written as
\begin{equation}
	\frac{\partial\left(\rho H\right)}{\partial t}+\nabla\cdot\left(\rho H\mathbf{v}\right)=\nabla\cdot k_{T}\nabla\theta.
\end{equation}
For above standard convection-diffusion equation, the evolution equation of the lattice Boltzmann model reads
\begin{equation}
	g_i\left(\mathbf{x}+\mathbf{c}_i\Delta t,t+\Delta t\right)=g_i\left(\mathbf{x},t\right)-\Lambda_{ij}^{\theta}\left(g_j-g_j^{eq}\right)\left(\mathbf{x},t\right)+\Delta t\left(\delta_{ij}-\frac{\Lambda_{ij}^{\theta}}{2}\right)G_j^{\theta}\left(\mathbf{x},t\right),
\end{equation}
where the distribution functions are given by
\begin{equation}
	g_i^{eq}=\begin{cases}
		\rho H+\left(\omega_i-1\right)\theta, & i=0,\\
		\omega_i\theta+\omega_i\mathbf{c}_i\cdot\rho H\mathbf{v}/c_s^2, & i\neq0,
	\end{cases}\quad G_i^{\theta}=\frac{\omega_i\mathbf{c}\cdot\partial_t\left(\rho H\mathbf{v}\right)}{c_s^2}.
\end{equation}
In addition, the macroscopic temperature is calculated by
\begin{equation}
	H=\frac{1}{\rho}\sum_ig_i,\quad \theta=\frac{H}{c_p}+\frac{c_p^s}{c_p}h\left(\psi\right),
\end{equation}
and the relaxation parameter $s_1^{\theta}$ can be given by $k_T=\left(1/s_1^{\theta}-1/2\right)c_s^2\Delta t$.

Finally, we would also like to point out that in the LBM, the collision matrix $\bm{\Lambda}=\left(\Lambda_{ij}\right)$ can be written as
\begin{equation}
	\bm{\Lambda}=\mathbf{M}^{-1}\mathbf{S}\mathbf{M}=\mathbf{M}_0^{-1}\mathbf{S}_0\mathbf{M}_0,
\end{equation}
where $\mathbf{M}=\mathbf{C}_d\mathbf{M}_0$ is the transformation matrix, $\mathbf{S}$ is the diagonal relaxation matrix, and $\mathbf{C}_d$ is a diagonal matrix formed by the powers of lattice speed $\hat{c}$. $\mathbf{S}_0=\mathbf{C}_d^{-1}\mathbf{S}\mathbf{C}_d$ is the new relaxation matrix. 
In the simulations of two-dimensional problems, the D2Q5 lattice structure is applied for the anisotropic Allen-Cahn equation as well as the temperature and solute equations, while for the conservative Allen-Cahn equation and Navier-Stokes equations, the D2Q9 lattice structure is adopted.

\bibliographystyle{elsarticle-num} 
\bibliography{references}
\end{document}